\documentstyle[11pt]{article}
\def\PRL #1 #2 #3{{\sl Phys.  Rev.  Lett.} {\bf#1} (#2) #3}
\def\NPB #1 #2 #3{{\sl Nucl.  Phys.} {\bf B #1} (#2) #3}
\def\NPBFS #1 #2 #3 #4{{\sl Nucl.  Phys.} {\bf B #2} [FS#1] (#3) #4}
\def\CMP #1 #2 #3{{\sl Commun.  Math.  Phys.} {\bf #1} (#2) #3}
\def\PRD #1 #2 #3{{\sl Phys.  Rev.} {\bf D #1} (#2) #3}
\def\PLA #1 #2 #3{{\sl Phys.  Lett.} {\bf #1A} (#2) #3}
\def\PLB #1 #2 #3{{\sl Phys.  Lett.} {\bf B #1} (#2) #3}
\def\JMP #1 #2 #3{{\sl J.  Math.  Phys.} {\bf #1} (#2) #3}
\def\PTP #1 #2 #3{{\sl Prog.  Theor.  Phys.} {\bf #1} (#2) #3}
\def\SPTP #1 #2 #3{{\sl Suppl.  Prog.  Theor.  Phys.} {\bf #1} (#2) #3}
\def\AoP #1 #2 #3{{\sl Ann.  of Phys.} {\bf #1} (#2) #3}
\def\PNAS #1 #2 #3{{\sl Proc.  Natl.  Acad.  Sci.  USA} {\bf #1} (#2) #3}
\def\RMP #1 #2 #3{{\sl Rev.  Mod.  Phys.} {\bf #1} (#2) #3}
\def\PR #1 #2 #3{{\sl Phys.  Reports} {\bf #1} (#2) #3}
\def\AoM #1 #2 #3{{\sl Ann.  of Math.} {\bf #1} (#2) #3}
\def\UMN #1 #2 #3{{\sl Usp.  Mat.  Nauk} {\bf #1} (#2) #3}
\def\FAP #1 #2 #3{{\sl Funkt.  Anal.  Prilozheniya} {\bf #1} (#2) #3}
\def\FAaIA #1 #2 #3{{\sl Functional Analysis and Its Application} {\bf
#1} (#2) #3}
\def\BAMS #1 #2 #3{{\sl Bull.  Am.  Math.  Soc.} {\bf #1} (#2)
#3} \def\TAMS #1 #2 #3{{\sl Trans.  Am.  Math.  Soc.} {\bf #1} (#2) #3}
\def\InvM #1 #2 #3{{\sl Invent.  Math.} {\bf #1} (#2) #3}
\def\LMP #1 #2 #3{{\sl Letters in Math.  Phys.} {\bf #1} (#2) #3}
\def\IJMPA #1 #2 #3{{\sl Int.  J.  Mod.  Phys.} {\bf A #1} (#2) #3}
\def\AdM #1 #2 #3{{\sl Advances in Math.} {\bf #1} (#2) #3}
\def\RMaP #1 #2 #3{{\sl Reports on Math.  Phys.} {\bf #1} (#2) #3}
\def\IJM #1 #2 #3{{\sl Ill.  J.  Math.} {\bf #1} (#2) #3}
\def\APP #1 #2 #3{{\sl Acta Phys.  Polon.} {\bf #1} (#2) #3}
\def\TMP #1 #2 #3{{\sl Theor.  Mat.  Phys.} {\bf #1} (#2) #3}
\def\JPA #1 #2 #3{{\sl J.  Physics} {\bf A#1} (#2) #3}
\def\JSM #1 #2 #3{{\sl J.  Soviet Math.} {\bf #1} (#2) #3}
\def\MPLA #1 #2 #3{{\sl Mod.  Phys.  Lett.} {\bf A #1} (#2) #3}
\def\JETP #1 #2 #3{{\sl Sov.  Phys.  JETP} {\bf #1} (#2) #3}
\def\JETPL #1 #2 #3{{\sl Sov.  Phys.  JETP Lett.} {\bf #1} (#2) #3}
\def\PHSA #1 #2 #3{{\sl Physica} {\bf A #1} (#2) #3}
\def\CQG #1 #2 #3{{\sl Class.  Quantum Grav.} {\bf #1} (#2) #3}
\def\SJNP #1 #2 #3{{\sl Sov. J.  Nucl. Phys. (Yadern.Fiz.)} {\bf #1} (#2) #3}
\def\a{\alpha}\def\b{\beta}\def\g{\gamma}\def\d{\delta}\def\e{\epsilon}

\def\k{\kappa}\def\L{\Lambda}\def\s{\sigma}\def\S{\Sigma}
\def\Th{\Theta}\def\Om{\Omega}\def\G{\Gamma}

\newcommand{\p}[1]{(\ref{#1})}
\begin{document}
\renewcommand{\thefootnote}{\fnsymbol{footnote}}
\thispagestyle{empty}
\begin{flushright}
{\bf
TUW 98--07 \\
hep-th/9802032 \\
1998, February 5}\\
Minor corrections \\
1998, February 25
\end{flushright}

\begin{center}
{\LARGE
 D=10 Dirichlet super--9--brane.}
\footnote{ Supported  in  part  by  the
INTAS Grants {\bf 96-308, 93-127-ext, 93-493-ext}
and by the Austrian Science Foundation under the Project {\bf P--10221}}

\vspace{1.0cm}
\renewcommand{\thefootnote}{\dagger} \vspace{0.2cm}

\medskip

{\bf Vladimir Akulov, Igor  Bandos$^1$, Wolfgang Kummer$^2$ \\
 and Vladimir  Zima$^3$ }

\medskip

{\it ${}^1$  Institute for Theoretical Physics, \\
NSC Kharkov Institute of Physics and Technology, \\
 310108, Kharkov,  Ukraine\\
 e-mail: bandos@kipt.kharkov.ua,
 bandos@tph32.tuwien.ac.at \\
 ${}^2$ Institut f\"{u}r Theoretische Physik, \\ Technische
 Universit\"{a}t Wien, \\ Wiedner Hauptstrasse 8-10, A-1040 Wien\\
 e-mail: wkummer@tph.tuwien.ac.at \\
{\it ${}^3$ Kharkov State University, \\
 310077, Kharkov,  Ukraine }
\date{January 1998}

}
\vspace{0.5cm}

{\bf Abstract}

\end{center}

\medskip

{\small
Superfield equations of motion for $D=10$ $type~IIB$
Dirichlet super-9-brane are obtained from the generalized action principle.

The geometric equations containing {\sl fermionic} superembedding equations
and constraints on the generalized field strength of Abelian gauge
field are separated from the proper dynamical equations and are found
to  contain   these dynamical equations among their consequences.

The set of superfield equations thus obtained
involves a $Spin(1,9)$ group valued superfield $h_\a^{~\b}$ whose leading
component appears in the recently obtained simplified expression
for the $\kappa$--symmetry projector of the D9-brane.
The Cayley image of this superfield coincides (on the mass shell) with
the field strength tensor of the world volume
gauge field characteristic for the Dirichlet brane.

The superfield description of the
super-9-brane obtained in this manner is known to be, on the one hand,
the nonlinear
(Born-Infeld) generalization of supersymmetric Yang--Mills theory and,
on the other hand, the theory of partial spontaneous breaking of
$D=10$ $N=IIB$ supersymmetry down to $D=10$ $N=1$.}

\medskip

\setcounter{page}1
\renewcommand{\thefootnote}{\arabic{footnote}} \setcounter{footnote}0

\newpage

\section{Introduction}

The $D=10$ $type~II$ superbranes (Dirichlet superbranes or super-Dp-branes,
where $p=d-1$ is a number of space-like dimensions of the brane world
volume) \cite{Dbranes}-- \cite{baku2}
at present are the object of
attention  due to exceptional role
they play in understanding of string dualities, in particular
in the approach related to the Matrix model \cite{DbraneM}.

Covariant and explicitly $\kappa$--symmetric  actions
for the super-Dp-branes were  obtained recently \cite{c1,schw,bt,schw1} and
used intensively for studying brane intersections
\cite{kallosh,kallosh2} as well as for an approach to quantize
superbranes \cite{kallosh1}. The progress in the latter becomes
possible because, as found in \cite{schw1}, the super-Dp-brane
admits the covariant gauge fixing conditions for $\kappa$-symmetry
\footnote{These means that, e.g. super-D0-branes are an analog of massive
superparticle with extended supersymmetry \cite{JL}. It was known that such
superparticles has $\kappa$--symmetry when the central charge
$Z$ (it carries
due to the presence of some 1-dimensional Wess--Zumino term in the action)
coincides with the particle mass $m$. The possibility of covariant gauge fixing
for $m \not= 0$ was proved and extensively  used in \cite{JL}. }.
This property was used in \cite{kallosh,kallosh1,kallosh2}
to revise the $\kappa$-symmetry
description and to present it in a more simple way opening  the
possibility for many applications.

 Even before the covariant action have been constructed, the
equations of motion for super--Dp-branes written in terms of
world volume superfields
had been obtained \cite{hs2}
(in the linearized approximation)
in the frame of superembedding approach.

This approach
was elaborated
for $D=10$ superstrings and $D=11$ supermembrane (super-M2-brane)
in \cite{bpstv,bsv} in the course of development of the doubly
supersymmetric geometric approach.
The latter can be regarded as the
supersymmetric generalization of the classical surface theory of the 19th
century (see e.g. \cite{Ei}) and describes  superbranes in terms of
extrinsic geometry of the world volume superspace of the brane embedded
 into the target superspace.

 The number of fermionic 'directions' of
the world volume superspace are considered to be half the fermionic
coordinates of  target superspace.  In such a case the local world volume
supersymmetry replaces the $\kappa$--symmetry of ordinary Green--Schwarz
formulation, and can be reduced to the kappa symmetry when the auxiliary
fields are excluded \cite{stv}--\cite{bsezgin}.

The two approaches to the super-Dp-branes
were united in \cite{bst}, where the generalized action
principle  \cite{bsv} has been applied to  the generic case of
super-D-p-branes,  and the general
form of the superfield equations (including  embedding equations and proper
dynamical equations in superfield form) has been obtained from the generalized
action functional.

The power of the superembedding approach can be seen in the history of the
effective description of the $M$- theory 5-brane \cite{5-brane}
as well. Again first the superfield equations
had been obtained in Ref. \cite{hs2}
before the covariant
action was found in Refs. \cite{blnpst,schw5}.

Subsequently, the equivalence of the equations of
motion following from  the superembedding superfield equations  with the
ones obtained from the covariant action \cite{blnpst,schw5} was proved
in \cite{5-equiv}. Since then both the covariant action
and superembedding equations
have found quite a number of applications
\cite{SezginWest,Sezgin25,SchwWrap,SorT}.

The  superembedding equations
for the generic case of super-D-p-branes has  been
studied in a linearized approximation in Ref. \cite{hs1}
(see also a brief
description of results for super-$D3$-brane in \cite{bst}).
As noted in
\cite{hs1,Sezginpriv} (see also \cite{c12}),
the basic superembedding equation (the so--called geometrodynamical
condition) is not enough to produce all the dynamical equations of
super-p-branes with higher values of $p$ completely.
The constraints for the world volume gauge superfield have to be
introduced as additional geometric equations for these cases.


In distinction to the 'classical' superembedding approach based on the
geometrodynamic equation only,
the generalized action approach \cite{bsv}  developed for the case of
super-Dp-branes in \cite{bst} produces all the equations (superembedding
ones, gauge field 'constraints' as well as the proper dynamical
equations) in superfield form.  Then the whole set of equations may be
split into the geometrical and proper dynamical part and  their
interrelations may be studied.

Here we realize  such program for the case of
$D=10$
$type~IIB$ super-$D9$-brane
\footnote{
While this paper was being written an alternative way to obtain
the complete set of geometric equations has been proposed
\cite{hsc}. It consists in the consideration of the fundamental strings
whose ends lie in the world volume superspace of D-branes.}.

It should be noted that the super-9-brane
in any case requires a separate study as
there are no bosonic directions orthogonal to the world volume
and thus the geometrodynamic equation \cite{stv,gs92,tp93,bsezgin},
being the basis of the superembedding approach in its
generic form \cite{bpstv,hs1,hs2,bpst,HSWAsp}, becomes trivial.
And thus the problem of superembedding description of super-9-brane
remained open till now.

Moreover, the 9--brane assumes a special role in the D-brane scan.
With covariant gauge fixing for the $\kappa$--symmetry it reduces
to supersymmetric generalization of
the $D=10$ Born-Infeld model and, as Wess-Zumino terms of all super-Dp-branes
vanish in the covariant gauge, all the gauge fixed super-Dp-brane actions can be obtained from the super-D9-brane one by
world--volume dimensional reduction \cite{schw1}.

On the other hand, this theory can be regarded as one of the embedding of
$D=10$, $N=1$ superspace into the $D=10$, $N=2$, $type~IIB$ superspace.
Hence the super-9-brane is the model of partial spontaneous breaking of
$D=10$ $N=2$ supersymmetry \cite{kallosh2}.

The partial supersymmetry  breaking in $D=4,6$ attracted much attention
recently \cite{Taylor,Galperin}.
The general theorem about the impossibility of such partial breaking were
overcome for the first time in ref. \cite{Polch}, where it was demonstrated that
it does not hold in the presence of brane solitons in supersymmetric
field theory.
It was then recognized  that any super-p-brane model describes
in particular the partial supersymmetry breaking (see e.g. \cite{BSTann}).
An approach to superfield description of superbranes based on the
partial supersymmetry breaking concept was proposed and elaborated
for $D=2$ superparticle and $D=4$ superstring in \cite{IK90}.

In
\cite{Galperin}
partial
supersymmetry breaking
in $D=4$ was studied
in the 'classical'  framework of nonlinear realizations \cite{volkov}.
Different multiplets were found there
to be of use for description of Goldstone fermions  of the partial supersymmetry
breaking and their  superpartners.

From our point of view (see also \cite{kallosh2})
such a multiplicity originates in the fact that all the
possibilities presented in \cite{Galperin} must be related to different
compactifications of super-9-brane model down to $D=4, 6$ dimensions
with rigid  breaking of the corresponding number of both linearly and
nonlinearly realized supersymmetries.

Thus the superfield equations describing
the super--9--brane can be regarded as a kind
of 'master' model of partial supersymmetry breaking
\footnote{
Recently, the relations between superembedding approach and
nonlinear realization one was studied in \cite{c12} on the example of
$D=7$ super--5--brane. }.

These
superfield equations are obtained here
 from generalized action principle
for $D=10$ $type~IIB$ super--D9--brane \cite{bst}.

The set of geometric equations are extracted and separated from
and the proper dynamical ones.
These geometric equations contain {\sl fermionic}
superembedding equation
and constraints on the generalized field strength of the abelian
world volume gauge
superfield.

We investigate their integrability conditions
and find that the geometric equations contain all  the dynamical equations
among their consequences.
This is just the situation described in
\cite{Sezginpriv,c12}. The presence of gauge field constraints
among the necessary geometric equations
creates certain problems for the use of superembedding approach without a
dynamical basis.  However, as we demonstrate, these problems disappear
when the superembedding is based on the generalized action \cite{bsv,bst}.

The superfield equations  obtained from the generalized action
involve a $Spin(1,9)$ group valued superfield $h_\a^{~\b}$ whose leading
component appears in the recently obtained simplified expression
for the kappa--symmetry projector of D9-brane
\cite{kallosh,kallosh1,kallosh2}.

This superfield can be regarded as a 'nonlinear square root'
of the field strength of the world volume gauge
(super)field characteristic for the Dirichlet brane.
More precisely, this field strength coincides with Cayley image of
the $SO(1,9)$ valued matrix $k$ corresponding to the $Spin(1,9)$ valued
spin--tensor $h$ (on the mass shell).


\bigskip

Thus we obtain the superfield description of the
super-9-brane which is known to be, on the one hand, the nonlinear
(Born-Infeld) generalization of supersymmetric Yang--Mills theory and,
on the other hand, the theory of partial spontaneous breaking of
$D=10$ $N=2$, $type~IIB$ supersymmetry down to $D=10$ $N=1$.

\bigskip

The paper is organized as follows.

We conclude this section by description
of our basic notations and conventions.

\bigskip

In Section 2 we describe the generalized action functional \cite{bst} for the
super--9--brane in flat $D=10,~type$ $IIB$ superspace.
The peculiar features of the
super-9-brane, allowing
to write the generalized action without
use of Lorentz harmonic variables, are considered here as well as
in Appendix A.

\bigskip

Section 3 is devoted to the variation of the generalized action.
After a preliminary discussion how to  extract the information
about equations of motion and symmetries from the external derivative of
the generalized action, in Subsection 3.1. we reproduce the calculation of
the external derivative
of the Lagrangian form  Ref. \cite{bst}. We then  use it
in Subsection 3.3 to recall the justification of
the (superfield) $\kappa$--symmetry.
In Subsection 3.2 it serves for the derivation of the general form of
superfield equations of motion in terms of
superforms. All this material can be regarded as a specialization
for the super-9-brane case of the
consideration presented in \cite{bst} for generic case of super-Dp-branes.
This is necessary for  the next sections as, in particular,
it provides us with the expressions for basic variations of
different pieces of Lagrangian form and with the identity for the
$\kappa$--symmetry projector matrix $\bar{\Gamma}$,
being of extremely importance for
the following.

\bigskip

In section 4 we justify the antidiagonal form of the
$\kappa$-symmetry projector $\bar{\Gamma}$  in the completely Lorentz covariant
way and prove that the $16 \times 16$ matrix $h_\b^{~\a}$
(denoted by $e^{\hat{a}}$ in \cite{kallosh,kallosh1,kallosh2}),
determining
completely
this projector,
takes its values in the $Spin(1,9)$ group.
In the frame of 'standard' component formulation
\cite{c1,schw,c2,schw1,bt} this result was obtained
by use of the special gauge \cite{kallosh} and it found wide applications
\cite{kallosh1,kallosh2}.
We, however find it instructive to present the completely covariant way of the
 proof based on the identity for $\bar{\Gamma}$
\cite{bst} referred to in the previous section.

As a bonus, we get in such a way a {\sl covariant} form of the
relation between the Lorentz group valued spin--tensor field $h$
and the auxiliary antisymmetric tensor field $F_{ab}$.
The latter coincides
with the  components of the generalized field strength
${\cal F}$ of the world volume 1--form gauge field $A$ on the
mass shell.
We obtain a set of useful identities between
$h_\b^{~\a}$ and $F_{ab}$ fields as well as
between their derivatives which are used in the next sections.

\bigskip

In Section 5 we rewrite the external derivative of the Lagrangian form
using the $h_\b^{~\a}$ spin-tensor and obtain a simple form of fermionic
(super)field equations.

The set of essential superfield equations is analyzed in Section 6.
Splitting   the set of these equations of motion into
geometrical  and
proper dynamical ones,
we prove that the geometric equations contain dynamical ones among
their consequences.
This is done by studying of their integrability conditions
 as it was done in \cite{bpstv} for
$D=10$ superstrings and the $D=11$ supermembrane (M-theory 2-brane).

 As we obtain {\sl all} the dynamical equations of super-9-brane from
 geometrical ones,
we have found the minimal set of superfield equations describing
the super--9--brane model and, hence, the nonlinear generalization of
$D=10$ supersymmetric Yang--Mills (SYM) theory (i.e.
$D=10$ Goldstone SYM multiplet
which describes the partial spontaneous breaking of $D=10$ type $IIB$
supersymmetry).

\bigskip

  To clarify the multiplet structure, in Section 6 we study the geometric
  equations in the linearized approximation.
  A relation between the
  main field strength
$W^\a $
of $N=1$ SYM multiplet
and the Goldstone fermion superfield of partially breaking $type~IIB$
supersymmetry is noted here.

\bigskip

In Conclusion we discuss briefly the obtained results.
The interdependence of the Bianchi identities and the integrability
conditions for the fermionic superembedding equations
is studied in the Appendix B.

\bigskip

\subsection{Basic notations and conventions}

Our notations are close to the ones of
\cite{bt} up to normalization and choice of the metric signature
(mostly minus in our article).

The coordinates of flat target superspace are denoted by
\begin{equation}\label{SSPtg}
  Z^{{M}}  =
 (X^m, \hat{\Th}^{\hat{\mu}} )   \equiv
 (X^m, \Th^{1\mu}, \Th^{2\mu})   , \qquad
\end{equation}
$$
m=0,1,...,9 , \qquad \mu = 1,...,16 , \qquad {\hat{\mu}} = (I,\mu ),
\qquad I = 1,2 , $$
$E^A$ is the  supervielbein 1-form of the flat $D=10$
$type~IIB$ superspace
\begin{equation}\label{EA} E^A = (E^a,
\hat{E}^{\hat{\a}} ) \equiv (E^a, E^{1\a}, E^{2\a} ) \end{equation} $$
a=0,1,...,9 , \qquad \a = 1,...,16 , \qquad {\hat{\a}} = (I,\a ),
I= 1,2 .
$$
Here $E^a$ is the
bosonic vielbein
\begin{equation}\label{Ea}
E^a \equiv \Pi^{m} \d^a_{m}
\end{equation}
\begin{equation}\label{Pi}
\Pi^{{m}} = dX^{{m}}
- i d\Th^1 \s^{{m}} \Th^1 - i d\Th^2 \s^{{m}} \Th^2
\equiv   dX^{{m}} - i d\hat{\Th}
\left( \matrix{ \s^{{m}} & 0 \cr
0 &  \s^{{m}}  }\right)
\hat{\Th} =
\end{equation}
$$
\equiv   dX^{{m}} - i d\hat{\Th}^{\hat{\mu}}
\left( I  \otimes  \s^{{m}} \right)_{\hat{\mu}\hat{\nu}}
\hat{\Th}^{\hat{\nu}},
$$
and

 \begin{equation}\label{Eal}
E^{\hat{\a}} =
d\hat{\Th}^{\hat{\mu}} \d_{\hat{\mu}}^{~\hat{\a}}, \qquad
\Leftrightarrow
\qquad
E^{I\a} = d\hat{\Th}^{I{\mu}} \d_{{\mu}}^{~\a}, \qquad I=1,2
\end{equation}
$$
d\hat{\Th}^{\hat{\mu}} = (d\Th^{1\mu},d\Th^{2\mu})
$$
are fermionic (Grassmann) vielbein forms.

Then the expressions of the torsion forms ('torsion constraints')
of the {\sl flat}
$D=10$, $type~IIB$ tangent superspace
are
\begin{equation}\label{Ta}
T^a \equiv {\cal{D}} E^a = dE^a =
-i ( E^{1{\a}}\wedge E^{1{\b}} + E^{2{\a}}\wedge E^{2{\b}})
\s^a_{{\a}{\b}} \equiv
\end{equation}
$$
\equiv
 - i E^{\hat{\a}}\wedge E^{\hat{\b}}
\left(I \otimes \s^a
\right)_{\hat{\a}\hat{\b}}
= - i E^{\hat{\a}}\wedge E^{\hat{\b}}
\left( \matrix{ \s^a _{{\a}{\b}} & 0 \cr
0 &  \s^a _{{\a}{\b}}  }\right) ,
$$
$$
T^{\hat{\a}} \equiv {\cal{D}}E^{\hat{\a}} = dE^{\hat{\a}} = 0 . \qquad
$$

The target superspace supervielbein  $E^A$ \p{EA},\p{Ea},\p{Eal},
being coincident with
the natural supervielbein of
flat $D=10, ~type~IIB$ superspace
$\Pi^M \equiv (\Pi^m, d{\Th}^{\hat{\mu}})$,
can be used for the construction of the generalized action
{\sl only for the super-D9-brane}.

In the general case of super-Dp-branes with $p<9$ the Lorentz harmonic variables
(see \cite{bzst,bpstv} and refs. therein) shall be included  into
Eqs. \p{Ea}, \p{Eal} instead of Kronecker symbols.
They are necessary to adapt
the target space frame to the bosonic world volume (see \cite{bpstv,bsv}):
The pull backs of $D-p-1$ vielbein forms $E^i$ entering the set of
$D=10$ bosonic vielbein forms $E^a=(E^{\tilde{a}},E^i)$
must vanish on the mass shell,
while the pull backs of the remaining $(p+1)$ forms $E^{\tilde{a}}$
will give rise to the set of linearly independent forms which can be used as a
vielbein forms of world volume superspace.
Such an adaptation appears dynamically as a result of variation with respect to
the harmonic variables \cite{bsv,bst}.

For the $9$--brane case where  there are no  bosonic directions orthogonal
to the world volume
we do not need harmonics to adapt the
bosonic frame and this is why
the generalized action
can be written without harmonics at all.

However we keep the separate notation for the
flat target space supervielbein to obtain some simplification of the equations
as well as to make
a connection  clear
between  our equations and ones
from Refs. \cite{bst,hs1,hs2,5-equiv}, where the branes in curved supergravity
background were considered.

For  the same reason, we keep  the covariant derivative symbol
${\cal D}$ as well in all those places where it should appear in the curved
$D=10$ $type~IIB$ superspace and/or for the lower dimensional branes,
although for the super--9--brane in the flat $D=10$ $type~IIB$ superspace
all the induced connections
\cite{bpstv} are trivial and
vanish for the natural choice of the supervielbein fixed by
 \p{EA}, \p{Ea}, \p{Eal}. Hence ${\cal D}=d$.

The basic volume form written in terms of the vielbeine is denoted by
\begin{equation}\label{E10}
 E^{\wedge 10} \equiv {1 \over (10)!} \e_{a_1...a_{10}}
E^{a_1} \wedge ... \wedge E^{a_{10}}
\end{equation}
$$
\equiv  {1 \over (10)!} \e_{m_1...m_{10}}
\Pi^{m_1} \wedge ... \wedge \Pi^{m_{10}} .
$$

The 'standard' bosonic $8$--form and $9$--form are normalized as
$$
E^{\wedge 8}_{ab} \equiv {1 \over 2 ^. 8!} \e_{aba_1...a_{8}}
E^{a_1} \wedge ... \wedge E^{a_{8}} ,
$$
$$
E^{\wedge 9}_{~a} \equiv {1 \over 9!} \e_{aa_1...a_{9}}
E^{a_1} \wedge ... \wedge E^{a_{9}} .
$$

The  list of products includes following useful identities
$$
E^{\wedge 9}_{~a} \wedge E^b = - \d_a^b E^{\wedge 10}, \qquad
$$

$$
E^{\wedge 8}_{ab} \wedge E^c = - \d_{[a}^c E^{\wedge 9}_{~b]}. \qquad
$$

The world--volume superspace of the super-D9-brane and its local coordinates
are denoted by
\begin{equation}\label{SSPwv}
{\S}^{(10|16)}= \{ z^M \} = \{ (\xi^m, \eta^\mu =\eta^\mu (\xi^m)) \}, \qquad
m=0,1,...,9 , \qquad \mu = 1,...,16
\end{equation}
and the intrinsic supervielbein forms on the world volume are
\begin{equation}\label{eA}
e^A = (e^a, e^\a ) \equiv dz^M e_M^{~A}, \qquad
a=0,1,...,9 , \qquad \a = 1,...,16  .
\end{equation}

The pull--backs of target space supervielbein forms onto the world volume
superspace are denoted by the same symbols
$E^{a}$, $E^{1\a}$, $E^{2\a}$.

\section{Generalized action functional for $D=10$ $type~IIB$
super-D9-brane}

The super-D-p-brane generalized action
\cite{bst} for the super-9-brane
in $D=10~type~IIB$ superspace
acquires the form

\begin{equation}\label{SLLL}
S = \int_{{\cal{M}}^{10}} {\cal{L}} =
 \int_{{\cal{M}}^{10}} ({\cal{L}}_0 +  {\cal{L}}_1 +  {\cal{L}}_{WZ})
\end{equation}
where
\begin{equation}\label{L0}
{\cal{L}}_{0} =  E^{\wedge 10} \sqrt{-det(\eta_{ab}+F_{ab})}  ,
\end{equation}

\begin{equation}\label{L1}
{\cal{L}}_{1} = Q_{8} \wedge (dA-B_2  - {1 \over 2} E^{a} \wedge E^{b} F_{ba})
\end{equation}
and the Wess-Zumino Lagrangian form is the same the one as
appearing in the by now standard formulation \cite{c1,schw,c2,schw1,bt}
\begin{equation}\label{LWZ}
{\cal{L}}_{WZ} = e^{\cal{F}} \wedge  C \vert_{10} ,
\qquad C = \oplus _{n=0}^{5} C_{2n} , \qquad
e^{{\cal F}}= \oplus _{n=0}^{5} {1\over n!} {\cal F}^{\wedge n}
\end{equation}
where the formal sum of the RR superforms $C= C_0 + C_2 +...$
and
of the powers of two form
${\cal F}$
is used and $\vert_{10}$ means the restriction to the $10$--superform input
\cite{c1,c2,bt}.

\bigskip

The two form ${\cal F}$ is
\begin{equation}\label{calF}
{\cal F} = dA - B_2
\end{equation}
where $B_2$ is the NS-NS two-form (super-)field whose field strength
is just the external derivative of the Wess--Zumino term of the
$type~IIB$ Green-Schwarz superstring
\begin{equation}\label{H3}
H_{3} \equiv dB_2 = i
E^{a} \wedge (E^{1{\a}}\wedge E^{1{\b}} -E^{2{\a}}\wedge E^{2{\b}})
(\s_a)_{\a\b}
\end{equation}
$$
\equiv i
E^{a} \wedge E^{\hat{\a}}\wedge E^{\hat{\b}}
\left(
\s_3 \otimes \s_a
\right)_{\hat{\a}\hat{\b}} \equiv
i E^{a} \wedge E^{\hat{\a}}\wedge E^{\hat{\b}}
\left( \matrix{ (\s_{a})_{\a\b} & 0\cr
0 & - (\s_{a})_{\a\b}  \cr }\right) .
$$

The 'vacuum' values of the RR superform gauge fields $C_{2n}$ is known to
be nonvanishing for  the case when  super-Dp-branes are present.
These 'vacuum' values are expressed in terms of target superspace coordinates
  $X$ and $\Theta$ only. However, explicit expressions
(see \cite{Dcan}) are rather complicated.
Fortunately we really need the curvatures
of RR superfields only, whose 'vacuum' values are much simpler.
It is convenient to write them using
the formal sum notations as well \cite{c1}--\cite{bt}
\begin{equation}\label{R}
R = e^{-{\cal{F}}} \wedge d ( e^{\cal{F}} \wedge C ) =
\oplus _{n=0}^{5} R_{2n+1}, \qquad
\end{equation}

\begin{equation}\label{R2n}
R_{2n+1} = {i \over (2n+1)!}
E^{a_{2n+1}} \wedge ... \wedge E^{a_1} \wedge E^{\hat{\a}}\wedge E^{\hat{\b}}
\left( \matrix{ 0 & (\s_{a_1...a_{2n+1}})_{\a\b} \cr
(-1)^n (\s_{a_1...a_{2n+1}})_{\b\a} & 0 \cr }\right)
\end{equation}
$$
= {2i \over (2n+1)!}
E^{a_{2n+1}} \wedge ... \wedge E^{a_1} \wedge E^{1{\a}}\wedge E^{2{\b}}
(\s_{a_1...a_{2n+1}})_{\a\b}
$$
where the symmetry properties of $D=10$ sigma matrices are used
to obtain the second  line.

$F_{ab}$ as included into
Eqs. \p{L0}, \p{L1} is an auxiliary antisymmetric tensor superfield.
The Lagrange multiplier term \p{L1} provides the
identification of the (pure bosonic) 2-form

\begin{equation}\label{Fform}
F \equiv {1 \over 2} E^b \wedge E^a F_{ab}
 \end{equation}
being constructed from this auxiliary field and the pull--backs of the
target space bosonic vielbeine \p{Ea}
with the generalized field strength \p{calF}
of the world volume abelian  gauge superfield $A=dz^M A_M (z)$
\begin{equation}\label{deltaQ}
{\d S \over \d Q_8} =0 \qquad \Rightarrow
F\equiv {1 \over 2} E^b \wedge E^a F_{ab}~=~{\cal F} \equiv dA - B_2 .
\end{equation}

\bigskip

The integration in \p{SLLL} is performed over an arbitrary
bosonic surface
$$
{\cal M}= (\xi^m, \eta^\mu =\eta^\mu (\xi^m)) \equiv z^M (\xi^m)
$$
in the world volume superspace
\p{SSPwv}.

Henceforth, the coordinate fields
entering \p{SLLL} shall be regarded
as world volume superfields taken on the ten dimensional bosonic surface
${\cal M}$
 \begin{equation}\label{superfieldsM}
X^m = X^m (\xi^m, \eta^\mu (\xi^m)), \qquad \Theta^{1\mu} =
\Theta^{1\mu} (\xi^m, \eta^\mu (\xi^m)) \qquad \Theta^{2\mu} =
\Theta^{2\mu} (\xi^m, \eta^\mu (\xi^m))
 \end{equation}
$$
A_M = A_M (\xi^m, \eta^\mu (\xi^m)), \qquad A= dz^M A_M =
d\xi^m (A_m + \partial_m \eta^\mu A_\mu) ,
$$
$$
F_{ab} = F_{ab} (\xi^m, \eta^\mu (\xi^m)), \qquad
$$
$$
Q_8 = {1 \over 8!} dz^{M_8}(\xi^m)\wedge ... \wedge dz^{M_1}(\xi^m)
Q_{M_1...M_8} (\xi^m, \eta^\mu (\xi^m)) .
$$

Thus the generalized action \p{SLLL} can be treated as one included
additional $16$ fermionic fields $\eta^\mu (\xi^m)$ in a quite nonlinear manner.

To arrive at  the equations of motion
one  varies with respect to these fermionic
fields (i.e. with respect to ${\cal M}$) on the same
footing as with respect to $X$ and $\Theta$.

Further
consideration of the
properties of the generalized action
can be found in \cite{bsv}
for the superbrane case and (in much more detail)
in \cite{rheo} for the case of supergravity
\footnote{In supergravity this approach is known under
the names 'group manifold' or
'rheonomic' one. Really it is  much more complicated than
our ('rheotropic' \cite{bsv}) approach to superbranes,
as the  basic objects for supergravity actions
 are curvature two forms instead of supervielbeine in the superbrane case.}

The key points to obtain the superfield equations  are

\begin{itemize}
\item It can be proved \cite{bsv} that, for the functional of the type
\p{SLLL} (i.e. written in terms of differential forms without the use
of the Hodge operation) the variation with respect to the surface
does not produce independent equations, i.e. the corresponding equations
are satisfied identically after the 'field' equations,
appearing as a result of variations with respect to
proper filed variables $X$, $\Theta$,... are taken into account
(see \cite{rheo,bsv} for details).

In this sense the generalized action is independent of the integration surface
${\cal M}$ and thus possesses a superdiffeomorphysm invariance (see
\cite{rheo}).

\item As a  result of the arbitrariness of the surface ${\cal M}$
and of the independence of the generalized action on this surface,
all the remaining equations, appearing as a result of the generalized action
variation with respect to 'proper' field variables ($X$, $\Theta$ etc.),
can be treated as superfield equations,
i.e. the equations for the world volume superfields and superforms
  \begin{equation}\label{superfields}
X^m = X^m (\xi^m, \eta^\mu ), \qquad \Theta^{1\mu} =
\Theta^{1\mu} (\xi^m, \eta^\mu ) \qquad \Theta^{2\mu} =
\Theta^{2\mu} (\xi^m, \eta^\mu )
 \end{equation}
$$
 A= dz^M A_M(\xi , \eta ), \qquad
F_{ab} = F_{ab} (\xi^m, \eta^\mu ), \qquad
$$
$$
Q_8 = {1 \over 8!} dz^{M_8}\wedge ... \wedge dz^{M_1}
Q_{M_1...M_8} (\xi^m, \eta^\mu ) ,
$$
which are not restricted to any surface \cite{bsv}.

\item Assuming the surface ${\cal M}$ to be the pure bosonic world volume
${\cal M}_0$
$$
{\cal M}={\cal M}_0= (\xi^m, \eta^\mu =0)
$$
and, thus, reducing  all the superfields to their leading components
  \begin{equation}\label{fields}
X^m = X^m (\xi^m, 0), \qquad \Theta^{1\mu} =
\Theta^{1\mu} (\xi^m, 0 ) \qquad \Theta^{2\mu} =
\Theta^{2\mu} (\xi^m,0 ),
 \end{equation}
we get a component formulation of super-D9-brane.
Such a formulation is   equivalent to the standard
(Dirac--Born--Infeld--like) one \cite{c1}--\cite{bt} (see  \cite{bst}
for the proof in generic case of super-Dp-branes) and possesses the
$\kappa$--symmetry, {\sl but in the irreducible form}.

\item For the generalized action the transformations giving rise to the
$\kappa$--symmetry on the component level (superfield $\kappa$--symmetry)
can be regarded as target space
or fiber representation of the superdiffeomorphysm invariance (see
\cite{bsv} and refs. therein).

\end{itemize}

\bigskip

\section{Variation of the generalized action.}

As the action is written in terms of differential forms,
one can extract the variation  from the external derivative
of the Lagrangian form using

 \begin{equation}\label{dLidL}
\d {\cal{L}} = i_\d d{\cal{L}} + d (i_\d {\cal{L}}) .
 \end{equation}
Then the last term can be dropped for the closed brane.

The application of \p{dLidL} to variations
which are not related to superdiffeomorphisms  requires  care regarding the
definition of the contractions of forms.
E.g. to vary the Wess--Zumino term with respect to the $A_M$ superfield
(being involved into $\d {\cal{L}}$ through its generalized field strength
${\cal F}\equiv dA - B_2$, $A=dz^M A_M$) we have to define
\begin{equation}\label{deltaA}
i_\d {\cal F} = \d A , \qquad i_\d C = 0, \qquad i_\d A = 0, \qquad i_\d E^A = 0 .
\end{equation}

In this way we get, e.g.
(cf. with eq. \p{0dLWZ} below)
$$
\d{\cal{L}}_{WZ} = e^{\cal{F}} \wedge  R\vert_{9} \wedge \d A
 = i E^{\hat{\a}}\wedge E^{\hat{\b}}\wedge
(\hat{\g})_{\hat{\a}\hat{\b}} \wedge e^{\cal{F}} \vert_{9} \wedge \d A
$$

Another important variation is the one for the fermionic coordinate fields
$\Theta^{\hat{\a}}= (\Theta^{1{\a}}, \Theta^{2{\a}})$.
It can be obtained from \p{dLidL} by supposing
\begin{equation}\label{deltaTh0}
i_\d {\cal F} = \d A - i_\d B_2 = 0, \qquad i_\d C = 0, \qquad i_\d A = 0, \qquad i_\d E^a = 0
\end{equation}
$$
i_\d E^{\hat{\a}} = \d \Theta^{\hat{\a}} .
$$

\subsection{External derivative of Lagrangian 10-form.}

Here we review the  calculation of the external derivative of the Lagrangian
form
\cite{bst}.

Using  \p{R}, \p{R2n}
the derivative of the Wess--Zumino term \p{LWZ} becomes

 \begin{equation}\label{0dLWZ}
d{\cal{L}}_{WZ} = e^{\cal{F}} \wedge  R\vert_{11} \equiv
 i E^{\hat{\a}}\wedge E^{\hat{\b}}\wedge
(\hat{\g})_{\hat{\a}\hat{\b}} \wedge e^{\cal{F}}
\end{equation}
with

 \begin{equation}\label{hatg}
(\hat{\g})_{\hat{\a}\hat{\b}} \equiv \oplus _{n=0}^{4}
\left( \matrix{ 0 & (-1)^n \hat{\s}^{(2n+1)} \cr
 \hat{\s}^{(2n+1)} & 0  \cr }\right)
\equiv \oplus _{n=0}^{4} (  \s_1  (\s_3)^n  \otimes \hat{\s}^{(2n+1)})
\qquad
\end{equation}

 \begin{equation}\label{hats}
\hat{\s}^{(2n+1)}\equiv {1 \over (2n+1)!}
E^{a_1} \wedge ... \wedge E^{a_{2n+1}}
\s_{a_1...a_{2n+1}} .
\end{equation}

It is convenient
to write \p{0dLWZ} as

 \begin{equation}\label{1dLWZ}
d{\cal{L}}_{WZ} =
 i E^{\hat{\a}}\wedge E^{\hat{\b}}\wedge
(\hat{\g})_{\hat{\a}\hat{\b}} \wedge e^F
+ ({\cal{F}}- F) \wedge S,
\end{equation}
where $S$ is defined by

 \begin{equation}\label{Sdef}
({\cal{F}}- F) \wedge S \equiv
(e^{\cal{F}}-e^F) \wedge  R =
(e^{\cal{F}}-e^F) \wedge i E^{\hat{\a}}\wedge E^{\hat{\b}}\wedge
(\hat{\g})_{\hat{\a}\hat{\b}}, \qquad
\end{equation}
or, formally,

\begin{equation}\label{Sdef1}
S = {e^{\cal{F}}-e^F \over {\cal{F}}- F } \wedge i E^{\hat{\a}}\wedge
E^{\hat{\b}}\wedge (\hat{\g})_{\hat{\a}\hat{\b}}. \qquad
\end{equation}
Here and below the restriction to the 11-form input
in the formal sums is assumed but not
written explicitly.

The derivative of the 'kinetic term'
\begin{equation}\label{0dL0}
d{\cal{L}}_{0} =  E^{\wedge 9}_a \wedge
{\cal D} E^a  \sqrt{-det(\eta_{ab}+F_{ab})}
+  E^{\wedge 10 }\wedge d\sqrt{-det(\eta_{ab}+F_{ab})}
\end{equation}
can be transformed by algebraic manipulations into
\cite{bst}
 \begin{equation}\label{1dL0}
d{\cal{L}}_{0} =  iE^{\wedge 9}_a \wedge
E^{\hat{\a}}\wedge E^{\hat{\b}}
 (I \otimes \s_b )_{\hat{\a}\hat{\g}}
\left( (\eta + \s_3 F)^{-1~ba} \otimes I \right)^{\hat{\g }}_{~\hat{\b }}
  \sqrt{-det(\eta_{ab}+F_{ab})} -
\end{equation}
$$
- E^{\wedge 8}_{ab} ~~(\eta + F)^{-1~ab} \wedge d({\cal F} - F)
 \sqrt{-det(\eta_{ab}+F_{ab})}   .
$$

Then, using the property $\bar{\G }^2=I$ of the $\kappa$--symmetry
projector matrix  \cite{c1}-\cite{bt}
\begin{equation}\label{Gbar}
\bar{\Gamma}_{\hat{\a}}^{~\hat{\b}} =
{1 \over \sqrt{-det(\eta +F)}}
 \S_{n=0}^{5} {(-1)^n\over 2^n{}^. n!}
F_{a_1b_1}...F_{a_nb_n} ((\s_3)^n \otimes \s^{a_1b_1...a_nb_n})^.
(-\s_1 \otimes I)
\end{equation}
(rewritten  in our case in terms of the auxiliary tensor)
and the
identity \cite{bst}
\footnote{This can be proved from \p{Gbar} by direct application
of the gamma matrices multiplication rule
$$
\s^{a_1b_1...a_nb_n}\s^c =
\s^{a_1b_1...a_n b_n c } + 2n \s^{[a_1b_1...a_n}\eta^{b_n]c}
$$}

\begin{equation}\label{IdGs}
E^{\wedge 9}_a ~~\bar{\G } ~(I \otimes \s_b )
\left( (\eta + \s_3 F)^{-1~ba} \otimes I \right) =
{1\over \sqrt{-det(\eta_{ab}+F_{ab})}} e^F \wedge \hat{\g} \vert_9
\end{equation}
we can represent  \p{1dL0} as
 \begin{equation}\label{2dL0}
d{\cal{L}}_{0} =  i
E^{\hat{\a}}\wedge E^{\hat{\b}}\wedge (\bar{\G }\hat{\g})_{\hat{\a}\hat{\b}}
\wedge e^F -
\end{equation}
$$
- E^{\wedge 8}_{ab}~(\eta + F)^{-1~ab} \wedge d({\cal F} - F)
 \sqrt{-det(\eta_{ab}+F_{ab})} .
$$
The first term in \p{2dL0} coincides with one of Eq. \p{1dLWZ} up to the
$\bar{\G}$. Thus we get

 \begin{equation}\label{dL0dLWZ}
d{\cal{L}}_{0}+d{\cal{L}}_{WZ} =  i
E^{\hat{\a}}\wedge E^{\hat{\b}}\wedge ((1+\bar{\G })\hat{\g})_{\hat{\a}\hat{\b}}
\wedge e^F -
\end{equation}
$$
- E^{\wedge 8}_{ab}~(\eta + F)^{-1~ab} \wedge d({\cal F} - F)
 \sqrt{-det(\eta_{ab}+F_{ab})}+ ({\cal{F}}- F) \wedge S .
$$

The external derivative
of the ${\cal{L}}_{1}$ becomes
\begin{equation}\label{dL1}
d{\cal{L}}_{1} =
dQ_8 \wedge ({\cal{F}}-F) + Q_8 \wedge d({\cal{F}}-F) \equiv
\end{equation}
$$
\equiv dQ_8 \wedge ({\cal{F}}-F) - Q_8 \wedge \Big( H_3 +
E^{b} \wedge T^a F_{ab}+ {1\over 2}E^{b} \wedge
E^a \wedge{\cal{D}} F_{ab}\Big) .
$$

\bigskip

Collecting  Eqs. \p{dL0dLWZ} and \p{dL1} we get

\bigskip

 \begin{equation}\label{dL}
d{\cal{L}} \equiv d{\cal{L}}_{0}+ d{\cal{L}}_{1}+d{\cal{L}}_{WZ} =
\end{equation}
$$
=
i E^{\hat{\a}}\wedge E^{\hat{\b}}\wedge ((1+\bar{\G })\hat{\g})_{\hat{\a}\hat{\b}}
\wedge e^F -
$$
$$
(dQ_8 + S) \wedge (dA-B_2 - {1 \over 2} E^b \wedge E^a F_{ab})
$$
$$
- \left(Q_8 - E^{\wedge 8}_{ab}(\eta + F)^{-1~ab}
\sqrt{-det(\eta+F)}\right)\wedge \Big( H_3 +
E^{b} \wedge T^a F_{ab}+ {1\over 2}E^{b} \wedge
E^a \wedge{\cal{D}} F_{ab}\Big) .
$$

\bigskip

\subsection{Bosonic superfield equations
following from the generalized action}

To obtain the bosonic superfield equations of motion
we consider contractions of \p{dL} with
a variation symbol obeying
$$
i_\d E^{\hat{\a}} =0   .
$$

Keeping the only nonvanishing contraction  in \p{dLidL}  to be
$i_\d dQ_8 \equiv \d Q_8$,
we get (cf. Eq. \p{deltaQ})

\begin{equation}\label{deltaQ1}
{\d S  \over \d Q_8}
\equiv {i_\d d{\cal L} \over \d Q_8} = 0 \qquad \Rightarrow
\qquad F\equiv {1 \over 2} E^b \wedge E^a F_{ab}~
=~{\cal F} \equiv dA - B_2 .
\end{equation}

If the only nonvanishing contraction is $i_\d {\cal D}F_{ab}=
\d F_{ab}$ we obtain from \p{dL} the expression for the
Lagrange multiplier $Q_8$
 \begin{equation}\label{Q8eq}
Q_8 = E^{\wedge 8}_{ab}(\eta + F)^{-1~ab}
\sqrt{-det(\eta+F)}  .
\end{equation}

The equation corresponding to the variation  of the world volume gauge superfield
appears for the  choice of only nonvanishing contraction to be
$i_\d dA=\d A$ (cf.  \p{deltaA}).

Then the only contribution comes  from
the second line in \p{dL} producing
\begin{equation}\label{deltaA1}
{\d S \over \d A} \equiv
{i_\d d{\cal L} \over i_\d {\cal F}} =0 \qquad \Rightarrow \qquad
dQ_8 = S ,
\end{equation}
where the 9-superform $S$ is defined by \p{Sdef}.
Substituting \p{Q8eq} into \p{deltaA1} we arrive at
the supersymmetric generalization of the Born-Infeld equation
\begin{equation}\label{BIsusy}
d(  E^{\wedge 8}_{ab}(\eta + F)^{-1~ab} ) =
{e^{\cal{F}}-e^F \over {\cal{F}}- F } \wedge i E^{\hat{\a}}\wedge
E^{\hat{\b}}\wedge (\hat{\g})_{\hat{\a}\hat{\b}} \vert_{ {\cal{F}}= F }~,
\qquad
\end{equation}
where the formal way of writing the contribution from the WZ term
\p{Sdef1} and Eq. \p{deltaQ1} are  used.

Only the first line in \p{dL} provides
 inputs into other variations
after
Eqs. \p{deltaA1}, \p{deltaQ1}, \p{Q8eq}
 are taken into account.

So, supposing the only nonvanishing contraction
to be $\d
X^{m} \equiv i_\d E^{a} \d_a^{~m}$, we obtain

\begin{equation}\label{deltaX}
{\d S \over \d X^{{m}}} =0 \qquad \Rightarrow \qquad
E^{\hat{\a}}\wedge E^{\hat{\b}}\wedge
i_{E^a} \Big[\Big((1+\bar{\G })\hat{\g}\Big)_{\hat{\a}\hat{\b}}
\wedge e^F \Big] = 0 ,
\end{equation}
which is satisfied identically due to the fermionic equations
of motion to be considered in detail in the next section.

Such a dependence of the equations obtained by variation  of the
$X^m$ variables is really the Noether identity reflecting the
(bosonic) general coordinate
invariance of the generalized action.
This bosonic general coordinate invariance together with
the superfield generalization of the $\kappa$--symmetry
form the superdiffeomorphysm invariance of the generalized action.

\subsection{On  fermionic equations and
$\kappa$--symmetry of super--9--brane}

The fermionic equations
\begin{equation}\label{deltaTh}
{\d S \over \d \Theta^{\hat{\mu}}}\equiv
{i_\d d{\cal L} \over i_\d E^{\hat{\mu}}}=0 ,
\qquad \Rightarrow \qquad
 E^{\hat{\b}}\wedge ((1+\bar{\G })\hat{\g})_{(\hat{\a}\hat{\b})}
\wedge e^F = 0
\end{equation}
appear for  $i_\d E^a =0$, $i_\d E^{\hat \a} = \d \hat{\Theta}^{\hat \a}
\not= 0$ from the contraction of first line of Eq. \p{dL}.

Before turning to a more careful investigation of them
let us note that they as well as the first line of
\p{dL} itself,  can be used
to prove the $\kappa$--symmetry
of the super-D9-brane actions at the component level,
as  done in \cite{bst} for the generic case of super-Dp-brane.

The $\kappa$--symmetry is provided by the fact that
the projector is present in the fermionic equations
\p{deltaTh},
and, thus
 only $16$  of $32$  fermionic forms
$E^{\hat{\a}}=(E^{1\a}, E^{2\a})$ are involved
into the derivative of the Lagrangian form as well as in the fermionic equations.

This
means that only half of them are independent. This  is just the
Noether identity reflecting the fermionic gauge symmetry of
the action. When the integration manifold
${\cal M}$ is chosen to be a pure bosonic world volume
 ${\cal M}_0$  (i.e. $\eta^\a = 0$), this fermionic gauge symmetry
is just the $\kappa$--symmetry of the component (Lorentz harmonic)
formulation.
For the generalized action such a symmetry also holds and can be regarded as
a tangent space representation of the superdiffeomorphysm invariance
(see \cite{bsv} and refs. therein).

\section{$\bar{\Gamma}$ and Lorentz group valued spin-tensor
variables $h_{{\a}}^{~{\b}}$ .}

The fermionic superfield equation \p{deltaTh}
are complicated. Thus it is extremely important to
find  variables which provide a significant simplification
of  \p{deltaTh}.

We demonstrate here that such  variables  really exist and
are related to the $\kappa$--symmetry projector matrix $\bar{\G}$.

\bigskip

As first emphasized in \cite{kallosh} and used intensively in
\cite{kallosh1,kallosh2},
the $32 \times 32$ projector matrix of the $\kappa$-symmetry transformations
$\bar{\Gamma}$ \p{Gbar} has the block--antidiagonal form
\begin{equation}\label{Ggauge}
\bar{\Gamma}_{\hat{\a}}^{~\hat{\b}} =
- \left(\matrix{ 0 &  h_{{\a}}^{~{\b}}
\cr
                   (h^{-1})_{{\a}}^{~{\b}} & 0 \cr }
\right) .
\end{equation}

Here we present a new covariant derivation of this (for the 9-brane case,
although the same can be done for the general case of any value of $p$)
as well as of the fact that the matrix $h$ takes its values in the
fundamental representation of $Spin(1,9)$ group.

Our constructive proof provides as a bonus a
{\sl covariant expression for $h$ in terms of
antisymmetric tensor $F_{ab}$}.
These results will be extremely important in the study of
the superfield equations of super-9-brane.

\subsection{ Antidiagonal form of $\bar{\Gamma}$,
Lorentz group valuedness of
$h$ superfield and its relation with $F_{ab}$}

This fact can be proved easily. Indeed,
in accordance with \p{IdGs}, multiplication by
$\bar{\Gamma}$ transforms the {\sl diagonal} matrix 9-form
$$
E^{\wedge 9}_a  \left( (\eta + \s_3 F)^{-1~ba} \otimes  \s_b \right)
= E^{\wedge 9}_a \left( \matrix{\s_b (\eta + F)^{-1~ba} & 0 \cr
 0 & \s_b (\eta - F)^{-1~ba} \cr }\right)
$$
into ${\hat{\g}}\wedge e^F$, the (block--)off--diagonal one \p{hatg}.
This forces the $\bar{\Gamma}$ to have off-diagonal.
Then the requirement $\bar{\Gamma}^2=1$ results in the condition for
off--diagonal blocks to be inverse matrices. Thus one gets \p{Ggauge}
\footnote{To prove the $\kappa$--symmetry
 we really need only in  the identity \p{IdGs} the part which is
 symmetric in the spinor indices. But also the statement above can be proved even
from the symmetric part only.
Indeed, e.g. for the 1-st left $16\times 16$ block
$A_\a^{~\b}$ of the $\bar{\Gamma}$ matrix we then
get $(A\s_b)_{\{ \a\b \} }  (\eta + F)^{-1~ba}= 0$.
Decomposing the $A$ matrix into the complete basis
 ($256=1+45+210$)
$$A_\a^{~\b}= A_0 \d_\a^{~\b} +
~ A^{ab}(\s_{ab})_\a^{~\b}+ A^{abcd}(\s_{abcd})_\a^{~\b}$$
and using the gamma matrix algebra
we find from $(A\s_b)_{\{ \a\b \} } = 0$
that all the irreducible parts of the $A$ matrix vanish
$$ A_0 =0, \qquad
 A^{ab}=0, \qquad A^{abcd}=0, $$
 and, hence,
$A_{\a}^{~\b } = 0$. Nevertheless, these details are not
necessary as  \p{IdGs} is true  in general
(i.e. not only for its symmetric part), see footnote 5.}.

From the identity \p{IdGs} for the $\bar{\Gamma}$ matrix
\p{Ggauge} we find
\begin{equation}\label{h1}
\sqrt{...} E^{\wedge 9}_{~a}
(h\s_b)_{\a\b} (\eta - F)^{-1~ba} = e^F
\wedge \oplus_{n=0}^4 (-)^n\hat{\s}^{2n+1}_{\a\b} \vert_9 , \end{equation}
\begin{equation}\label{h-1}
\sqrt{...} E^{\wedge 9}_{~a}
 (h^{-1}\s_b)_{\a\b} (\eta + F)^{-1~ba} = e^F \wedge \oplus_{n=0}^4
\hat{\s}^{2n+1}_{\a\b} \vert_9 ,
\end{equation}
where  (cf. \p{hats})
$$
e^F \wedge \oplus_{n=0}^4 \hat{\s}^{2n+1}_{\a\b} \vert_9 =
E^{\wedge 9 }_c ~\S_{n=0}^4 {(-1)^n\over 2^n n!} F_{a_1b_1}...   F_{a_nb_n}
\s^{a_1b_1... a_nb_nc}_{\a\b} ,
$$
$$
e^F \wedge \oplus_{n=0}^4 (-)^n \hat{\s}^{2n+1}_{\a\b} \vert_9 =
E^{\wedge 9 }_c ~\S_{n=0}^4 { 1 \over 2^n n!} F_{a_1b_1}...   F_{a_nb_n}
\s^{a_1b_1... a_nb_nc}_{\a\b} .
$$

Now one can find that in $D=10$ the $16 \times 16$
matrix valued forms $\hat{\s}^{2n+1}_{\a\b}$ entering \p{h1}, \p{h-1}
have the symmetry properties
\footnote{In $D=10$ the complete basis for the matrices with two lower
Majorana--Weyl spinor indices is provided by the  symmetric matrices
$\s_a, ~\s_{a_1...a_5}$ and the antisymmetric ones $\s_{a_1a_2a_3}~~$
($ 16 \times 16 = 256 = 136 + 120 = 10 + 126 + 120$).}

$$
\hat{\s}^{2n+1}_{\a\b}= (-1)^n \hat{\s}^{2n+1}_{\b\a} .
$$
In this way one finds that the expression
\p{h1}  for $(h\s_b)_{\a\b}(\eta - F)^{-1~ba}$
coincides with \p{h-1} taken for $(h^{-1}\s_b)_{\b\a}\equiv
(\s_bh^{-1})_{\a\b}$

\begin{equation}\label{hId0}
(h\s_b)_{\a\b} (\eta - F )^{-1~ba} =
(h^{-1}\s_b)_{\b\a}(\eta + F)^{-1~ba} .
\end{equation}

An equivalent form of Eq. \p{hId0} is
\begin{equation}\label{Id1}
(h\s^a h^T)_{\a\b} = \s^b_{\a\b} k_b^{~a}  \qquad
\end{equation}
with
\begin{equation}\label{Id2}
 k_b^{~a} = \d_b^{~a} - 2 ((\eta+F)^{-1}F) _b^{~a}
\equiv ((\eta+F)^{-1}(\eta-F))_b^{~a}
\equiv ((\eta-F)(\eta+F)^{-1})_b^{~a} .
\end{equation}

Eq. \p{Id2} is the Cayley construction for the pseudoorthogonal
('$\eta$-orthogonal' or Lorentz group valued) matrix $k$
$$
(k^T)_b^{~a} \equiv  k_{~b}^a =(k^{-1})_b^{~a}
\qquad \Leftrightarrow \qquad k_b^{~a} \qquad \in \qquad SO(1,9)
$$
for the antisymmetric Cayley image $F_{ab}$  (see e.g.
\cite{Postnikov}).

Hence \p{Id1} defines the spin-tensor field
$h_\b^{~\a}$ as taking its values
in the double covering of the Lorentz group $Spin(1,9)$:
\begin{equation}\label{hinS}
h_\b^{~\a} \qquad \in \qquad Spin(1,9) .
\end{equation}

For completeness let us present the expression for the contraction
of $h_\b^{~\a}$ spin tensors
with tilde sigma matrices, which
is expressed through the $k$ matrix as well  as it can be shown
\begin{equation}\label{Id3}
(h^T\tilde{\s}^a h)^{\a\b} = \tilde{\s}^{b~ \a\b} k_b^{~a} . \qquad
\end{equation}

\bigskip

\subsection{Derivatives of $h_\a^{~\b}$}

As $h_\b^{~\a}$ field takes the values in $Spin(1,9)$ group and
is one of the two element of this group ($\pm h$) corresponding
to the vector rotation $k$ \p{Id1},
the derivative of $h$ is related to the  derivative of $k$ by the standard
isomorphism relation between $spin(1,9)$ and $so(1,9)$ algebras
\footnote{This can be obtained straightforwardly by taking the derivative
of Eq. \p{Id1} and  by using the $D=10$ Fierz identities
(see \cite{bzst} for more details
concerning Lorentz group valued quantities).}
\begin{equation}\label{dhdF}
(h^{-1}dh)_\b^{~\a} = {1 \over 4} (k^{-1}dk)^{ab} ~ (\s_{ab})_\b^{~\a} .
\end{equation}
At the same time, the derivatives of $k$ can be calculated in terms of $F_{ab}$
directly from the expression \p{Id2}
\begin{equation}\label{dkFdFF}
dk^{ab} =
- 2 (\eta +F)^{-1~ac} dF_{cd}  (\eta +F)^{-1~db} .
\end{equation}
Thus the direct relation between the  derivatives of $h_\b^{~\a}$
 and ones of the $F_{ab}$ tensor is
\begin{equation}\label{dhFdFF}
dh_\b^{~\a} = - {1 \over 2} (\eta - F)^{-1~ac} dF_{cd}
(\eta +F)^{-1~db} (h\s_{ab})_\b^{~\a} .
\end{equation}

\section{ Spin-tensor $h_\b^{~\a}$ and fermionic equations of motion.}

Using the identity \p{IdGs}  inversely,  Eq. \p{dL} can be represented
in another equivalent way.
For the first line in \p{dL}, which is the only essential below,
we get

 \begin{equation}\label{1dL}
d{\cal{L}} =  i
E^{\hat{\a}}\wedge E^{\hat{\b}}\wedge
((1+\bar{\G })\hat{\g})_{\hat{\a}\hat{\b}}
+ ...
\equiv
\end{equation}
$$
\equiv
iE^{\wedge 9}_a \wedge
E^{\hat{\a}}\wedge E^{\hat{\b}} \left( (1+\bar{\G })^.
 (I \otimes \s_b )^.
 ((\eta + \s_3 F)^{-1~ba} \otimes I) \right)_{\hat{\a}\hat{\b }}
  \sqrt{-det(\eta_{ab}+F_{ab})}
+ ...
$$
As
$$
(I \otimes \s_b )^.  (\eta + \s_3 F)^{-1~ba} \otimes I
\equiv \left( \matrix{ \s_b  (\eta + F)^{-1~ba}  & 0 \cr
 0 & \s_b  (\eta - F)^{-1~ba}\cr }\right) , \qquad
$$

$$
(\eta - F)^{-1~ba}=(\eta + F)^{-1~ab}
$$
we have for the r.h.s. of \p{1dL}
 \begin{equation}\label{EPE}
E^{\hat{\a}}\wedge E^{\hat{\b}}\wedge ((1+\bar{\G })
(I \otimes \s_b )^.
 ((\eta + \s_3 F)^{-1~ba} \otimes I))_{\hat{\a}\hat{\b}}
=
\end{equation}
$$
=
(E^2 - E^1h) \wedge ( - h^{-1}\s_b  (\eta + F)^{-1~ba} E^1 +
\s_b  (\eta - F)^{-1~ba} E^2) .
$$
in the gauge \p{Ggauge}.

With \p{Id1}, \p{Id2}
we can
write
$
(h^{-1}\s^a) (\eta +F)^{-1}=
(\s^b h^T)(\eta -F)^{-1}{}_b^{~a} , \qquad
$
and further transform \p{EPE}  into
 \begin{equation}\label{EPE1}
E^{\hat{\a}}\wedge E^{\hat{\b}}\wedge ((1+\bar{\G })
(I \otimes \s_b )^.
 ((\eta + \s_3 F)^{-1~ba} \otimes I))
=  (E^2 - E^1h) \wedge
\s_b  (\eta - F)^{-1~ba} (E^2- E^1h) .
\end{equation}

\bigskip

Thus we get for \p{1dL}
 \begin{equation}\label{3dL0dLWZ}
d{\cal{L}} =
 i E^{\wedge 9}_a \wedge (E^2 - E^1h)^\a \wedge
(\s_b)_{\a\b} ~ (E^2- E^1h)^\b~  (\eta - F)^{-1~ba}\sqrt{-det(\eta_{ab}
+F_{ab})} + ...
\end{equation}
where
$...$ denotes the second and third lines in \p{dL}.

Hence,  the
fermionic superfield equations \p{deltaTh}
 following from the generalized
action
can be represented as (cf. with the D3--brane case \cite{bst,baku2}, where
the h-variables are defined, however, in a completely different way)
 \begin{equation}\label{Theq0}
E^{\wedge 9}_b \wedge ~(E^2 - E^1h)^\a
~(\s^a)_{\a\b} (\eta - F)^{-1~ab} = 0 .
\end{equation}
Decomposing \p{Theq0} into  basic forms
\footnote{Remember that the generic expression for any world volume
superform is e.g.
 $E^{I\a}\equiv e^\b E^{I\a}_\b + e^a E^{I\a}_a$ with some
set of $16$ independent Grassmann forms
$e^\b$ and
some intrinsic bosonic vielbein $e^a$ which will be identified with
the pull--backs of the forms $E^{1\a}$ and
$E^a$
below. Note that
the superfield equations for fermions \p{Theq0}  acquires the 'standard' form
$$
\tilde{e}^{\wedge 9}_a \wedge ~(E^2 - E^1h)^\a
~(\s^a)_{\a\b} = 0
$$
 similar to one for
the $type~I$ superbranes in terms of adequate variables
$\tilde{e}^a = 2E^{b} (\eta - F)^{-1}{}_b^{~a}$ (cf.
\cite{baku2} where, however, a completely different parametrization of the
projector matrix related to another variant of the induced fermionic vielbein of
the world volume superspace has been considered for the D3-brane).}
one gets
 \begin{equation}\label{Theq1}
E_\b^{2\a} - E_\b^{1\g}h_\g^{~\a} = 0,
\end{equation}
\begin{equation}\label{Theq2}
(\s_a)_{\a\b} (\eta - F)^{-1~ab}(E_b^{2\a} - E_b^{1\g}h_\g^{~\a}) = 0 .
\end{equation}
In terms of differential forms \p{Theq1} becomes
\begin{equation}\label{E2hE1}
E^{2\a} = E^{1\b} h_\b^{~\a} + E^a \psi_a^\a , \qquad
\end{equation}
whereas \p{Theq2} remains
\begin{equation}\label{dyneq0}
(\s_{a})_{\b\a} (\eta - F)^{-1~ab}\psi_b^\a
= 0, \qquad
\end{equation}
with
$
\psi_a^\a \equiv (E_a^{2\a} - E_a^{1\g}h_\g^{~\a})
$
by definition.

\section{Geometrical equations and proper equations of motion}

Eq. \p{E2hE1}
can be regarded as a  fermionic superembedding equation.
One more equation which can be treated as geometrical one appears
as a result of variation with respect to $Q_8$
\p{deltaQ1}.

Here we  prove that the geometric equations \p{E2hE1}, \p{deltaQ1}
 \begin{equation}\label{E2hE11}
E^{2\a} = E^{1\b} h_\b^{~\a} + E^a \psi_a^\a , \qquad
\end{equation}
\begin{equation}\label{dAF}
{\cal F} \equiv dA - B_2 = F , \qquad F \equiv {1 \over 2}
E^{a} \wedge E^{b} F_{ba} , \qquad
\end{equation}
following from
the generalized action contain  the
dynamical equations
\p{dyneq0}
\begin{equation}\label{dyneq}
(\s_{a})_{\b\a} (\eta - F)^{-1~ab}\psi_b^\a = 0, \qquad
\end{equation}
among their consequences.

To this end we study the integrability conditions
\begin{equation}\label{IEE}
I_2^\a \equiv
{1 \over 2} e^A \wedge e^B I_{BA}^\a
\equiv d(E^{2\a} - E^{1\b} h_\b^{~\a} - E^a \psi_a^\a)
= 0, \qquad
\end{equation}
\begin{equation}\label{JFF}
J_3 \equiv {1 \over 3!} e^A \wedge e^B \wedge e^C J_{CBA}
\equiv
d({\cal F} - F)= 0 . \qquad
\end{equation}
Eq. \p{JFF} is the superfield Bianchi identities for the world volume
gauge super-1-form.

\subsection{Induced world volume geometry}

\subsubsection{Induced supervielbein and 'supersymmetric static gauge'}

Eqs. \p{IEE}, \p{JFF} contain the decomposition of the
2--form and 3--form integrability conditions on the
arbitrary basis of the cotangent world volume superspace
$e^A = (e^a, e^{\a}) $.
However we find it convenient to use the world volume supervielbein
induced by embedding in the sense of the  identification
\begin{equation}\label{ea2}
e^a = E^a \equiv \Pi^m \d_m^a , \qquad
\end{equation}
\begin{equation}\label{eal2}
e^\a = E^{\a 1} \equiv d\Theta^{1\mu} \d_\mu^{~\a}.
\end{equation}

Eq. \p{eal2} provides the holonomic representation for
Grassmann vielbein forms. This reflects the possibility to identify
the Grassmann coordinate of the world volume superspace
$\eta^\mu$ with the coordinate function $\Theta^{1\mu}$:

\begin{equation}\label{Thid}
\Theta^{1\mu} = \eta^\mu .
\end{equation}
Such an identification (being a 'superpartner' of the so--called static gauge
widely used in the description of solitonic branes in supergravity
\cite{Stelle})
breaks the general supercoordinate invariance as an independent symmetry,
retaining however the invariance
under combined transformations including the same local shift
of $\Theta^{1\mu} $.  The possibility of the latter
transformations is provided by  the superfield extension of
the (irreducible) $\kappa$ symmetry which leaves invariant the generalized action.

Taking the leading component of Eq. \p{Thid}
$$
\Theta^{1\mu}\vert_{\eta^\mu =0} = 0
$$
we obtain the covariant gauge fixing  condition for the $\kappa$ symmetry
of the component formulation of  \cite{schw1,kallosh,kallosh1,kallosh2}.

Hence the identification \p{Thid}
can be regarded as a superfield generalization of such a
covariant gauge fixing.

\subsubsection{Covariant derivatives induced by embedding}

The  world volume covariant derivatives induced by the
embedding form the basis in tangent world volume superspace dual
to \p{ea2}, \p{eal2}

\begin{equation}\label{Dind}
d \equiv dz^M \partial_M =
e^a D_a + e^{\a} D_\a \equiv  E^a D_a + E^{1\a} D_\a  .
\end{equation}

For such a choice we find that the Grassmann field
$\psi_a^\a$ appearing in the
superembedding equation \p{E2hE11}
acquires the form of a vector covariant derivative of the
$\Theta^2$ supercoordinate function

\begin{equation}\label{psi2}
\psi_a^\a = E_a^{2\a} \equiv D_a \Theta^{2\a}  .
\end{equation}

The commutator of the induced covariant derivatives
$$
[D_A, D_B \} = - t_{AB}{}^C D_C
$$
is expressed in terms of the  component
of the induced torsion, which is basically
the pull--back of the target space torsion forms \p{Ta}.

Thus  expression for the torsion of flat target superspace
\p{Ta}
\begin{equation}\label{Ta2}
T^a \equiv dE^a = - i (E^{{\a}1}\wedge E^{{\b}1}+ E^{{\a}2}\wedge
E^{{\b}2}) \s^a _{{\a}{\b}}, \qquad T^{\a} \equiv dE^{1\a } = 0
\end{equation}
and the fermionic superembedding equations Eq. \p{E2hE11},
taken together with \p{Id1}, provide us with the following expressions for
the induced world volume torsion:
\begin{equation}\label{taa}
t^a \equiv
de^a = - 2 i e^{{\a}}\wedge e^{{\b}} \s^b _{{\a}{\b}} (\d + F
)^{-1}{}_b^{~a} + 2 i e^{b}\wedge e^{{\b}} (h\s^a \psi_b)_{{\b}} - i e^b
\wedge e^{c} ( \psi_b\s^a \psi_c), \qquad
\end{equation}
$$ t^{\a} \equiv de^{\a} \equiv dE^{1\a} = 0 . $$

Then the commutation relations between
the covariant derivatives induced by the embedding are
\begin{equation}\label{DDD}
\{ D_\a, D_\b \} =  4 i
\s^b _{{\a}{\b}} (\d + F )^{-1}{}_b^{~a} D_a ,
\qquad
\end{equation}
$$
[ D_\a, D_b ] = -2i  (h\s^a \psi_b)_{{\b}} D_a ,
\qquad
$$
$$
[ D_a, D_b ] = 2i  ( \psi_a\s^c \psi_b) D_c .
\qquad
$$

\subsection{Dynamical equations from   geometrical ones}

Using  \p{taa}
we find that the lowest dimensional
(in inverse length units) nontrivial components of the integrability
conditions \p{IEE}, \p{JFF} result respectively in the equations
\begin{equation}\label{I1}
I_{\b\g}^{~~\a} = 0 ,
\qquad \Rightarrow \qquad
D_{(\g} h_{\b ) }^{~\a} =  i (\s^{a})_{\b\g} (\d + k)_b^{~a} \psi_a^\a
= 2i (\s_b)_{\b\g} (\eta + F)^{-1}{}^{ba} \psi_a^\a ,
\qquad
\end{equation}
\begin{equation}\label{J1}
J_{\a\b c}=0 ,
\qquad \Rightarrow \qquad
(h\s^ah^T)_{\a\b} (\eta + F)_{ab} =
\s^a_{\a\b} (\eta - F)_{ab} .
\qquad
\end{equation}

It is remarkable that  Eq. \p{J1} coincides with  \p{Id1}, \p{Id2}
$$
(h\s^a h^T)_{\a\b} = \s^b_{\a\b} k_b^{~a},   \qquad
$$
$$
k_b^{~a} =
((\eta+F)^{-1}(\eta-F))_b^{~a}
\equiv ((\eta-F)(\eta+F)^{-1})_b^{~a}  \qquad \in \qquad SO(1,9) .
$$
Hence the Lorentz group valuedness of the spin tensor field
$h_{\b}^{~\a}$
$$
h_{\b}^{~\a} \qquad \in \qquad Spin(1,9)
$$
and the relation of this field with antisymmetric
tensor $F_{ab}$ are
contained in the integrability condition of geometric equations.

The restriction which \p{I1} puts on the $h$ superfield can be written
in  closed form as
\begin{equation}\label{I1hh}
\tilde{\s}_{a_1 ... a_5}{}^{\b \g}
D_{\g} h_{\b }^{~\a} = 0 .
\qquad
\end{equation}

The
components of the integrability conditions
\p{IEE}, \p{JFF},
carrying dimensions $3/2$ and $5/2$ respectively produce
\begin{equation}\label{I2}
I_{b\b}^{~~\a}=0 ,
\qquad \Rightarrow \qquad
D_{b} h_{\b}^{~\a} =
D_\b \psi_b^\a
+ 2i (h\s^{a}\psi_b)_{\b} \psi_a^\a  ,
\qquad
\end{equation}

\begin{equation}\label{J2}
J_{\a b c}=0 ,
\qquad \Rightarrow \qquad
D_{\a}F_{bc} = - 4i (h\s^a\psi_{[b})_{\a} (\eta - F)_{c]a} .
\qquad
\end{equation}

With \p{dhdF} and after some algebraic manipulations
\p{J2} becomes an expression for the Grassmann derivative  of the $h$ superfield
\begin{equation}\label{J2h}
D_{\b}h_\g^{~\a} = -2i (h\s^a\psi_{c})_{\b} (\eta + F)^{-1~cb}
(h\s_{ba})_\g{}^\a
\qquad
\end{equation}

From \p{J2h} we can get the expression for the symmetric part
$D_{(\g} h_{\b ) }^{~\a}$ of
the Grassmann derivative  of the
spin-tensor superfield $h$, which is already specified in terms of
$\psi$ and by Eq. \p{I1}.
Comparing these two equations yields

\begin{equation}\label{psi1}
 (h\s^a)_{(\b | \d}
(h\s_{ba})_{ |\g )}{}^\a
(\eta - F)^{-1~bc} \psi_{c}^\d
= -
  (\s_{b})_{\b\g} (\eta + F)^{-1}{}^{ba} \psi_a^\a
\qquad
\end{equation}
With \p{Id1}, \p{Id2} the r.h.s. of \p{psi1} can be written as
$$
 (h\s^{a}h^T)_{\b\g} (\eta - F)^{-1}{}_b^{~a} \psi_a^\a
$$
and thus  the same expression $(\eta - F)^{-1}{}_b^{~a} \psi_a^\a $
appears on  both sides of the equation.

Now it is easy to extract $h \times h$ factor
and thus to arrive at
\begin{equation}\label{psi22}
 (\s^a)_{(\b | \d}
(\s_{ab})_{ |\g )}{}^\a
(\eta - F)^{-1~bc} \psi_{c}^\d
=
  (\s_{b})_{\b\g} (\eta - F)^{-1}{}^{ba} \psi_a^\a .
\qquad
\end{equation}
Contracting  \p{psi22}  with $(\tilde{\s}^{a})^{\b\g}$
we obtain after straightforward algebra
$$
 (\tilde{\s}^{a})^{\b\g} {\s}_{b~\g\a}(\eta - F)^{-1}{}^{bc} \psi_c^\a = 0
$$
which evidently results in the superfield fermionic equations
of motion \p{dyneq0}
\begin{equation}\label{dyneq1}
(\s_{b})_{\b\a}(\eta - F)^{-1}{}^{bc} \psi_c^\a
= 0
\end{equation}
following from the generalized action.

\subsection{Other integrability conditions}

The higher dimensional integrability conditions, i.e. the ones with
all lower indices bosonic, are dependent in the sense that all their consequences
can be obtained by applying the covariant Grassmann derivative to
the equations \p{I1}, \p{J1}, \p{I2}, \p{J2}.
We prove this fact
in the Appendix B
using the
'identity for identity' technique (see \cite{zima} and refs therein, and
Appendix C in ref. \cite{bpstv}).

We, however, present the explicit form of the equations following from these
integrability conditions for completeness

\begin{equation}\label{I3}
I_{ab}^\a=0 ,
\qquad \Rightarrow \qquad
D_{[a} \psi_{b]}^{~\a} =  -
i (\psi_{[a}\s^{c} \psi_{b]}) \psi_c^\a ,
\qquad
\end{equation}

\begin{equation}\label{J3}
J_{abc}=0 ,
\qquad \Rightarrow \qquad
D_{[a } F_{bc]} = -2i( \eta + F)_{d[a}  (\psi_{b}\s^{d} \psi_{c]}) .
\qquad
\end{equation}

Eq. \p{J3} is just the bosonic Bianchi identity, but written in terms
of the induced covariant vector derivatives \p{Dind}.

\bigskip

Thus we have proved that the geometric equations \p{E2hE11}, \p{dAF} contain
{\sl all the
information about the super-9--brane theory}, and hence provide a
superfield description of
the nonlinear (Born--Infeld) generalization of the
$D=10$ SYM theory, included in the generic model of partial supersymmetry
breaking: from
$D=10, ~N=2, ~type~IIB$ to  $D=10,~N=1$.

 \section{Geometric equations in linearized approximation}

To make the multiplet structure of the super-D9-brane model more
transparent and the interrelation with
the consideration from Refs.  \cite{hs1,hsc} more explicit,
we reproduce here the investigation of the geometric equation
in the linearized approximation using the physical ('static') gauge
\footnote{Note that in the static gauge the close relation of the
superembedding approach \cite{bpstv,bsv,hs1} with the one \cite{IK90}
based on the partial supersymmetry breaking concept \cite{Polch} becomes
clear, though the first one still remains simpler \cite{c12}}.

\subsection{Physical gauge and linear approximation}

 The basic condition of the physical gauge is \p{Thid}
\begin{equation}\label{pg1}
\Theta^{1\mu}= \eta^{\mu} .
\qquad
\end{equation}
Taking into account the use of the linearized approximation in fields
in what follows,
it is convenient to introduce a
Grassmann Goldstone fermion superfield $W^{\mu}$ for partially broken
$type~IIB$ supersymmetry
as the differences between
two Grassmann coordinate superfields $\Theta^1$ and $\Theta^2$.
Hence in the gauge \p{pg1}
\begin{equation}\label{pg2}
\Theta^{2\mu}= \eta^{\mu} + W^{\mu}.
\qquad
\end{equation}
and we can write the bosonic gauge fixing conditions in the linearized
approximation as
\begin{equation}\label{pg3}
X^m + i \eta \s^m W = \xi^m
\qquad
\end{equation}

 To justify such choice, one
can consider the Bianchi identities
\p{JFF}
\begin{equation}\label{JFF0}
J_3 \equiv
d({\cal F} - F)= -H_3 - E^b \wedge T^a F_{ab}
- {1 \over 2} E^b \wedge E^a \wedge d F_{ab} =
\qquad
\end{equation}
$$
- i  E^b \wedge d\eta^{\a} \wedge d\eta^{\b} \s^a_{\a\b} (\eta - F)_{ab} +
 i  E^b \wedge d\Theta^{2\a} \wedge d\Theta^{2\b} \s^a_{\a\b} (\eta +
 F)_{ab} - {1 \over 2} E^b \wedge E^a \wedge d F_{ab} = 0.  $$
As the $0^{th}$ order term in $E^b$ should be the flat bosonic vielbein
$w^b$ (see below), it can be seen that one loses a consistency in
$0^{th}$ order input into the Eq. \p{JFF0}, when  assumes that
$\Theta^2$ is of the $1^{st}$ order.
On the other hand, \p{pg2} reflects the presence of the first term
with invertible spin tensor $h$ in the r.h.s. of the
fermionic superembedding equation
\p{E2hE11}.

With \p{pg1}--\p{pg3} we get for a bosonic vielbein in the linearized
approximation
\begin{equation}\label{Ealin}
E^a = w^a - 2i d\eta \s^a W .
\qquad
\end{equation}
Here
\begin{equation}\label{wa}
w^m = d\xi^m  - 2i d\eta \s^a \eta
\qquad
\end{equation}
is a bosonic vielbein of the flat world volume superspace.

The flat Grassmann derivatives appear in the decomposition
$$
d=
d\xi^m \partial_m +
d\eta^{\mu} \partial_{\mu} =
w^m \partial_m +
d\eta^{\mu} D_{\mu}
$$
and have the form
\begin{equation}\label{Dflat}
D_{\mu} =
\partial_{\mu} +
 2i ( \s^m \eta )_\mu
\partial_m .
\qquad
\end{equation}

The algebra of the flat derivatives is
$$
\{ D_\mu ,D_\nu \} = 4i \s^m_{\mu\nu} \partial_m .
$$

In this section we will denote the complete derivatives
by ${\cal D}_{\a}$.
Their expressions up to first order terms in fields are
$$
{\cal D}_{\a}= D_\a +2i (\s^m W)_\a \partial_m ,
$$
while for the bosonic derivatives one obtains
$$
{\cal D}_{a}= \d_a^m \partial_m .
$$

\subsection{Linearized geometric equations}

 With \p{pg1} -- \p{pg3} the linearized
the Bianchi identities
 \p{JFF0}
are
\begin{equation}\label{JFF01}
J_3 =
2 i  w^b \wedge d\eta^{\a} \wedge d\eta^{\b}
(\s^a_{\a\b}  F_{ab} +
2 D_{(\a | } W^\g \s_{b\g | \b )} -
\qquad
\end{equation}
$$
- {1 \over 2 }  w^b \wedge w^a \wedge d\eta^{\a}
(D_{\a}  F_{ab} + 4i \partial_{[a } W^\b \s_{b]\b \a})
- {1 \over 2 }  w^b \wedge w^a \wedge w^c
\partial_{[c } F_{ab]} .
$$

The most essential lowest dimensional component of
\p{JFF01} is
\begin{equation}\label{J1l}
J_{\a\b c}=0 ,
\qquad \Rightarrow \qquad
F_{ab} \s^b_{\a \b} =
2 D_{(\a | } W^\g \s_{a\g | \b )} .
\qquad
\end{equation}

Substituting the general decomposition for the
$DW$ superfield on the spinor indices
\begin{equation}\label{DWdec}
D_\a W^\b =
a^0 \d_\a^{~\b}
+a^{ab} \s_{ab}{}_\a^{~\b}
+a^{abcd} \s_{abcd}{}_\a^{~\b}
\end{equation}
one can find from \p{J1l} that the coefficients $a^0$ and $a^{abcd}$
vanish, i.e.
\begin{equation}\label{J1l10}
D_{\b } W^\b = 0, \qquad
\s^{abcd}{}_\b^{~\a} D_{\a } W^\b = 0, \qquad
\end{equation}
 and $a^{ab} = - {1 \over 4} F^{ab} $.
Thus Eq. \p{J1l} is equivalent to
\begin{equation}\label{J1l1}
D_{\a } W^\b
= - {1 \over 4} F_{ab} \s^{ab}{}_{\a}^{~ \b} .
\qquad
\end{equation}


Thus the fermionic superembedding equation becomes
 (cf. \p{pg2})
\begin{equation}\label{I1l}
dW^\a
= - {1 \over 4} d\eta^\b \s^{ab}{}_{\b}^{~ \a} F_{ab}
+ d\xi^m \partial_m W^\a .
\qquad
\end{equation}
This means that
\begin{equation}\label{hlpsil}
h_\b^{~\a} =
\d_\b^{~\a} - {1 \over 4} F_{ab} \s^{ab}{}_{\b}^{~ \a} , \qquad
\psi^\a_a = \d_a^m \partial_m W^\a
\qquad
\end{equation}
in the linear approximation.

\subsection{Linearized dynamical equations from the Bianchi identities}

 Acting on Eq. \p{J1l1} by Grassmann derivative and taking the symmetric
 part with respect to the lower spinor indices, one gets
\begin{equation}\label{J1l2}
- 8 i \s^m_{\a\b} \partial_m W^\g
= D_{(\a }F_{ab} \s^{ab}{}_{\b )}^{~ \g}  .
\qquad
\end{equation}
 Contracting \p{J1l2} with $\d_\g^\b$ we arrive at
\begin{equation}\label{J1l21}
- 16 i \s^m_{\a\b} \partial_m W^\b
=
\s^{ab}{}_{\a }^{~ \b}
D_{\b} F_{ab}.
\qquad
\end{equation}
On the other hand, \p{J1l2} means
\begin{equation}\label{J1l3}
D_{(\a }F_{ab} \s^{ab}{}_{\b )}^{~ \g}
=
{ 1 \over 16}
\s^m_{\a\b}
D_{\d }F_{ab} (\tilde{\s}_m\s^{ab})^{\d\g} .
\qquad
\end{equation}
Contracting \p{J1l3} with $\d_\g^{~\b}$
and using the gamma matrix algebra ($\tilde{\s}^m \s_{ab}\s_m =
6 \tilde{\s}_{ab} = - 6 \s_{ab}^T$) we obtain
\begin{equation}\label{J1leqF}
\s^{ab}{}_{\a}^{~ \b}
D_{\b} F_{ab} = 0
\qquad
\end{equation}
which is just the fermionic equation of motion
in the linearized approximation
(cf. with \p{J1l21})
\begin{equation}\label{J1leqW}
 \s^m_{\a\b} \partial_m W^\b = 0 .
\qquad
\end{equation}

To get the equations of motion for the gauge fields it is enough
to take a divergence from the basic equation
\p{J1l}
\begin{equation}\label{J1lv}
\partial^a
F_{ab} \s^b_{\a \b} =
2 D_{(\a  }
\s^m_{\b ) \g}
\partial_m
W^\g .
\qquad
\end{equation}
and use \p{J1leqW} to get
the standard Yang-Mills equations
\begin{equation}\label{J1leqFF}
\partial^a F_{ab}  = 0  .
\qquad
\end{equation}
The Bianchi identities with the solution $F_{ab} =
\partial_a A_b
- \partial_b A_a$ are included into Eq. \p{JFF01}.

\subsection{Comments}

Thus we have analyzed  the geometric equations
of the  super-D9-brane in the linearized approximation
and  obtained the dynamical equations
from the geometrical ones.

These results are not unsuspected as the
Eqs. \p{J1l}, \p{J1l10},  \p{J1l1} are characteristic for $D=10$ SYM theory
with the main field strength $W^{\a}$ related to the $dim~3/2$ component
$F_{\a b}$ of the complete field strength $F_{AB} = D_{A} A_{B} -
(-)^{AB} D_B A_A$
by $$ F_{\a b} \propto (\s_b W)_{\a } . $$ (The latter equation
follows from the superspace Bianchi identities for $F_{AB}$ restricted by
the constraints $F_{\a\b}=0$).

In this respect it is interesting to note that
the covariant
(with respect to $N=1$ supersymmetry and gauge symmetry) field strength
$W^\a $
of $N=1$ SYM multiplet originates in
  the Goldstone fermionic superfield (cf. \p{pg2})
 in the  linearized approximation.

\section{Conclusion and discussion}

In this paper we have obtained the superfield  equations
of motion for the $D=10,~type ~IIB$ Dirichlet super-9-brane
from the generalized action \cite{bst}.

We justify by a covariant proof
the statement from
refs. \cite{kallosh,kallosh1} concerning the antidiagonal form of
the $\kappa$--symmetry projector
$\bar{\G}$
as well as  the Lorentz group valuedness of the
spin--tensor (super)field $h_\b^{~\a}$ which determines the projector
$\bar{\G}$ completely.
For that we use only
 the identity for
$\bar{\G}$ \cite{bst}.

As our proof is constructive, we  obtain , as a bonus, a covariant
 relation between $h$ superfield and the auxiliary tensor
$F_{ab}$ coinciding with the field strength of the world volume gauge
field on mass shell. This relation means that $F_{ab}$ is the Cayley image
of a Lorentz group valued matrix $k$ whose double covering is provided by
the spin-tensor superfield $h$.

The derivation and investigation of the superfield equations of motion
are considerably simplified by using the
superfield $h$ and its covariant  relation with $F_{ab}$.

The prize to pay for this is the Lorentz group valuedness of the spin
tensor superfield $h$ which, however, does not create any problem
as the technique of dealing with Lorentz group valued variables have been
developed already in \cite{bzst}.

The set of geometric equations following from the generalized action
is extracted and separated
from the proper dynamical equations of motion.
Studying the integrability conditions for the first ones
we show as a consequence, that the geometric equations
\p{E2hE11}, \p{dAF}
contain {\sl all the
information about the super-9--brane theory} and, hence about
the nonlinear (Born--Infeld) generalization of the
$D=10$ SYM theory included in the generic model of partial supersymmetry
breaking.

The separate studying of the geometric equations in the linear
approximation displays an interesting point:
The Goldstone superfield $W^\mu = \Theta^{2\mu}-\Theta^{1\mu} $ of
the nonlinearly realized supersymmetry plays the role of the
field strength of the D=10 SYM multiplet in  this approximation
(cf. \p{J1l10}, \p{J1l1})
\footnote{When this paper was being written we received a copy of
\cite{hsc} which shows a certain  overlap with  the
Section 7. In particular, Eqs.
\p{J1l10}, \p{J1l1} can be extracted from that work as well.}.

Our results provide further evidence for a power of the
superembedding approach based on the generalized action \cite{bsv,bst} for
studying superbrane physics.

\bigskip

Another description of the super-9-brane
model  is provided by considering an
'extrinsic' geometry of
the world volume superspace.
This just corresponds to the approach of the
classical theory of surfaces (see \cite{Ei} and refs. therein),
which supersymmetric generalization  has been developed
in \cite{bpstv}.

The key observation is that the set of the integrability conditions
\p{IEE}, \p{JFF} together with the torsion constraints
\p{taa}  contain all the information
provided by  the original geometric equations
\p{E2hE11}, \p{dAF}.

In accordance with Eq. \p{I3}, \p{DDD}, the $\psi_a^\a$ superfield
should be regarded as  covariant derivative of some Grassmann
spinor superfield \p{psi2}. In such a way the
supercoordinate function
$\Theta^2$
appears in the extrinsic geometry.
At the same time, in this description
the superfield $\psi_a^\a$
is dependent because it is simply the notation for
a nonvanishing part of  the
Grassmann derivative of the $Spin(1,9)$ group valued superfield
$h_{\b }^{~\a}$ \p{I1}.

Thus the Lorentz group valued spin tensor superfield
$h_\a^{~\b}$  can be considered as the main superfield
for the description of the super-9-brane world volume superspace
'extrinsic' geometry.
The main equation which extract the super-D9-brane theory, or,
equivalently, $D=10$ Goldstone SYM multiplet from the $h_\a^{~\b} ~\in
~Spin(1,9)$ superfield
is \p{I1hh}
\begin{equation}\label{I1hh1}
\tilde{\sigma}_{a_1 ... a_5}{}^{\b \g}
D_{\g} h_{\b }^{~\a} = 0 .
\qquad
\end{equation}
The fermionic dynamical equations \p{dyneq} has the form
\begin{equation}\label{dyneqh}
D_{\a}h^{-1}{}_\b^{~\a} = 0 ,
\qquad
\end{equation}

Eq. \p{I1hh1} can be used in searching for a nonlinear generalization of
the action with infinitely many Lagrange multipliers superfield
proposed recently in \cite{BH}.

The development of the
'extrinsic' geometry description
(or doubly supersymmetric geometric
approach)
for the
super--D9--brane
is an interesting task for further study.
(Note that it is quite nonlinear as the
world volume geometry is characterized by the covariant derivative algebra
\p{DDD} involving the $h$ superfield).

\bigskip

A direct application of our results would be to study the
dimensional reduction
of our equations down to $D=4$ and
to arrive in this way at
the equations of the Goldstone  multiplets
found in \cite{Galperin}.
Another way  consists in constructing a generalized action
for $D=4,~N=2$ super-D3-brane, getting the  superfield equations
and  investigating their relations with ones from \cite{Galperin}.

 One more interesting problem for further study consists in the
 possibility to reformulate the generalized action in terms of $h$
 superfields instead of its Cayley image $F_{ab}$, as it was done
 for D3-brane in \cite{baku2}.
 Such investigation can provide an insight in searching for a generalized
 action for M-theory  super-5-brane (see \cite{bpst}) and, more generally,
 for unification of the superembedding approach with the PST technique
 \cite{pst} widely used for a Lagrangian description of  theories
 with self-dual gauge fields (see
 \cite{blnpst,schw5,SchwWrap,D10IIB,D11sd}).

\section{Acknowledgements}

 The authors are grateful to D. Sorokin, M. Tonin and E. Sezgin for
 useful discussions and comments.  The work  was supported in
 part by the INTAS Grants {\bf 96-308, 93-127-ext, 93-493-ext} and by the
 Austrian Science Foundation under the Project {\bf P--10221}.

\newpage

\section*{Appendix A: Remarks on harmonics and frames.}

In distinction with the generic super-D9-brane case \cite{bst},
the generalized action
for the super-9-brane
can be written
without spinor harmonics at all.

However, a reader can understand (if he like) all the equations as ones
including the Lorentz group valued Lorentz harmonic variables
\cite{bzst,bpstv,bsv} (and refs. in \cite{bpstv}).
as a bridge between target space and world indices. I.e. he can substitute
$$
E^a =
\Pi^m u^{~a}_m , \qquad E^{\a I} =
d\Theta^{I\mu} v_{\mu}^\a
$$
in any of our equations (except for ones from section 7).

If the harmonics are retained, the following points are important to remember
for the 9--brane case:
\begin{itemize}
\item
 As $~u^a_{{m}}~$
is the complete ($10 \times 10$) Lorentz group valued matrix,
we have the $SO(1,9)$ gauge invariance in the action
 \p{SLLL} and, thus, the
harmonic degrees of freedom are pure gauge ones.
\item
 The natural covariant derivative
(\cite{bzst,bpstv,bsv} and refs. therein)
\begin{equation}\label{Du}
{\cal D}  u^a_{{m}} \equiv
d u^a_{{m}}  - u^b_{{m}} \Om_b^{~a} = 0
\end{equation}
vanishes acting on the harmonics.
\item
 As the connection  $\Om_b^{~a}$
and, hence, the induced spin connection defined as pull back
of the connection form onto the world volume superspace
$~~\Om_b^{~a} = dz^M \Om_M{}_b^{~a}~~$
is $~~\Om^{ab}= u^{a{m}} d u^b_{{m}}~~$
\p{Du} and thus trivial, the induced covariant derivative
on the world volume superspace being defined as a pull--back of
\p{Du}
$$
{\cal D} = dz^M {\cal D}_M = e^A {\cal D}_A \qquad
$$
produces no curvature
 $$
 {\cal D}{\cal D}=0 . $$
\end{itemize}

 As there are no $E^i$ (i.e. bosonic directions orthogonal to the
9-brane in $D=10$ space-time), we do not need in harmonics to adapt the
bosonic frame to the world volume and we can use them to relate our results to
any frame.
However, the harmonics can be used to relate a general Lorentz
frame with the frame, where the  antisymmetric tensor $F$ (auxiliary in our case)
takes the form (cf. \cite{kallosh})

\begin{equation}\label{Fgauge}
F_{ab} =
\left(\matrix{ 0 &
\left( \matrix{\Lambda_0 & 0 & ... & 0 \cr
                   0 & \Lambda_1 & ... & 0 \cr
                    ... & ... & ... & ... \cr
                       0 & 0 & ... & \Lambda_4 \cr }
\right) \cr
\left( \matrix{- \Lambda_0 & 0 & ... & 0 \cr
                   0 & - \Lambda_1 & ... & 0 \cr
                    ... & ... & ... & ... \cr
                       0 & 0 & ... & - \Lambda_4 \cr }
\right)
& 0 \cr }
\right) ,
\end{equation}
i.e.
$$
  [F_{ab}] = i \s_2 \otimes diag(\Lambda_0, \Lambda_1, ..., \Lambda_4 ) .
$$

This gauge is very important in many respects
(see \cite{kallosh,kallosh1,kallosh2}).
So, our covariant relations for $h$ and $F$  could have been  obtained in
an alternative way by the
use of the gauge
\p{Fgauge} \cite{kallosh}, where
the spin-tensor field $h$ acquires the form
\begin{equation}\label{hgauge}
h_{{\a}}^{~{\b}}=
{\left((1-\L_0 \s^{05})(1-\L_1 \s^{16})...(1-\L_4 \s^{49})\right)_{{\a}}^{~{\b}}
\over
\sqrt{(1-\L_0^2)(1+\L_1^2)... (1+\L_4^2)}} .
\end{equation}

\bigskip

\section*{Appendix B: Interdependence of geometric equations}

Here we will study the interdependence of the integrability conditions
\p{IEE}, \p{JFF}.

It should be stressed that the fermionic superembedding equations
\p{E2hE11} considered with unrestricted $h_\a^{~\b}$
has a conventional character. Indeed, the $\psi$ superfield is
equal to the bosonic derivative of the $\Theta^2$ superfield
$$\psi_a^{~\a} = D_a \Theta^{2\mu} \d_\mu^{~\a} $$
by definition (cf. \p{dyneq0}). Thus the only
nontrivial statement containing in the equation \p{E2hE11}
with unrestricted $h$ is the
supposition about linear independence of the pull--backs of
$16$ fermionic forms $E^{1\a} = d\Theta^{1\mu} \d_\mu^{~\a}$, which
are used  as a basis in the fermionic sector of the world volume
superspace \p{eal2}.

On the other hand, as we have seen above, the dynamical equations
appear when the integrability conditions for  both geometric equations
\p{E2hE11} and \p{dAF} are taken into account.

\bigskip

\subsection*{B1. Interdependence of the higher dimensional integrability
conditions}


 We begin by studying  a dependence of some higher dimensional
 integrability conditions inside the sets \p{IEE} and \p{JFF}.

 To this end, following the line realized for the Bianchi identities
 of supergravity \cite{zima},
 we
 investigate  the integrability conditions for Eqs. \p{IEE} and \p{JFF}
 ('identities for identities' or
 'integrability conditions for integrability conditions')

\begin{equation}\label{I3EEE}
I_3^\a \equiv
{1 \over 3!} e^A \wedge e^B \wedge e^C I_{CBA}^\a
\equiv dI_2^\a = 0, \qquad
\end{equation}
\begin{equation}\label{J4FFF}
J_4 \equiv {1 \over 4!} e^A \wedge e^B \wedge e^C \wedge e^D  J_{DCBA}
\equiv dJ_3= 0 , \qquad
\end{equation}

Their components are
\begin{equation}\label{IABC}
{1 \over 3} I_{ABC}^{~~~~\a} \equiv
D_{\{ A } I_{BC\} }^{~~~~\a} +
t_{\{ A B| }^{~~~D} I_{D|C\} }^{~~~~\a} = 0 ,
\end{equation}
\begin{equation}\label{JABCD}
{1 \over 4} J_{ABCD} \equiv
D_{\{ A } J_{BCD\} } +
{3 \over 2} t_{\{ A B| }^{~~~E} J_{E|CD\} } = 0 .
\end{equation}

Supposing that all the integrability conditions \p{IEE} of
dimensions less then 2 are satisfied
$$
I_{\b \g  }^{~~~\a} = 0, \qquad
I_{\b c  }^{~~\a} = 0, \qquad
$$
we obtain from the component of \p{IABC} of dimension $2$
\begin{equation}\label{depI1}
t_{\b\g } ^{~~b}
I_{bc} ^{~~\a} = 0 . \qquad
\end{equation}
As for our choice of induced geometry \p{ea2}, \p{eal2}
the world volume torsion
is determined by Eq. \p{taa}
\begin{equation}\label{taa1}
t_{\b\g } ^{~~a}
=
\s^b_{\b\g } (\eta + F)^{-1}{}_b^{~a} , \qquad
\end{equation}
we get from \p{depI1}
$$
I_{bc} ^{~~\a} = 0 . \qquad
$$
This means that all the contents of the dim $2$ component of the
integrability conditions  \p{IEE} can be obtained from the corresponding
action by covariant Grassmann derivatives
on their components of dimensions $1$ and $3/2$. In other worlds,
these components of integrability condition \p{IEE} are dependent.

In the same manner, using the dim $3$ component $J_{\a\b cd}=0$ of
 Eq. \p{JABCD}, one can prove
 the dependence of the
integrability conditions $$ J_{bcd}=0 $$ of Eq. \p{JFF} on the low
dimensional components
$$ J_{\a \b c} = 0, \qquad J_{\a b c} = 0 $$
of the same equation
(remember that the dim $3/2$ component $J_{\a\b\g}=0$ of the  Eq. \p{JFF}
is satisfied identically for the case under consideration).

If one turns to the component $J_{\a\b\g d}=0$  of dim $5/2$ of the Eq.
\p{JABCD} and assumes that  the integrability conditions
\p{JFF} of dimension $2$, i.e.
$J_{\b\g c}=0$,
are satisfied identically,
one obtains
\begin{equation}\label{dep2}
\s^b_{\{ \a \b } \Psi_{\g \} ba} = 0 , \qquad  with \qquad
 \Psi_{\g ba} \equiv J_{\g cd} (\eta + F)^{-1}{}_b^{~c}
 (\eta + F)^{-1}{}_a^{~d} .
\end{equation}
(Here the expression \p{taa1} for the world volume torsion have been
used).

The general solution of Eq. \p{dep2}
(with respect to the lower indices)
is
 \begin{equation}\label{dep2s}
\Psi_{\g  ba} = {1 \over 10}
\s_{b\g\b}~
\tilde{\s}^{c \b \a}\Psi_{\a  ca}  \qquad
\end{equation}
(it can be obtained by contraction of \p{dep2} with
$\tilde{\s}^{b \b \g}$). Then, using the antisymmetry property of
$\Psi_{\g ba}$ \p{dep2} with respect to permutation of vector indices
$b,~a$
  \begin{equation}\label{dep2s1}
\Psi_{\g  ba} =
{1 \over 10}
\s_{b\g\b}~
\tilde{\s}^{c \b \a}\Psi_{\a ca}
\equiv
{1 \over 10}
\s_{a\g\b}~
\tilde{\s}^{c \b \a}\Psi_{\a bc} , \qquad
\end{equation}
and contracting \p{dep2s1} with $\tilde{\s}^{b \b \g}$,
we obtain
$$
\tilde{\s}^{c \b \a}\Psi_{\a bc} = 0, \qquad
$$
and, hence (cf. \p{dep2s})
$$
\Psi_{\a bc} = 0 \qquad  \Rightarrow \qquad J_{\a bc} = 0. \qquad
$$

Thus we have proved that only the lowest (dim $2$) component
$J_{\a\b c}=0$ \p{J1} of the integrability conditions
\p{JFF} for the geometric equation \p{dAF} (nonlinear SYM constraints)
{\sl can be independent}. I.e. all the results of the Eqs.  \p{J2}
(\p{J2h}) and \p{J3} can certainly be obtained by acting by the Grassmann
derivatives on the Eq.  \p{J1}.

\bigskip

In contradistinction, the independent set of
integrability conditions \p{IEE} can contain, in principle, not only the
lowest dimensional component $$ I_{\b\g}^{~~\a} = 0 $$ \p{I1}, but
some irreducible parts of Eq. \p{J2} as well.

 If one consider the {\sl dim} $3/2$ component $I_{\b c} ^{~~~\a}=0$
 of the Eq. \p{I3EEE}
 in assumption that $I_{\b\g}^{~~\a}=0$, one obtains
  \begin{equation}\label{dep3}
\s^b_{\{ \b\g|} \Psi_{b |\d \} }^{~\a } = 0
\end{equation}
 with
  \begin{equation}\label{redef1}
\Psi_{b\b}^{~~ \a} \equiv
(\eta - F)^{-1}{}_b^{~a} h^{-1}{}_\b^{~\g} I_{a\g}^{~~ \a}
\equiv
 (\eta - F)^{-1}{}_b^{~a}
 ( h^{-1}{}_\b^{~\g}  D_a h_\g^{~\a}
- h^{-1}{}_\b^{~\g} D_\g \psi_a^{\a} -
2i (\s^c\psi_a)_{\b} \psi_c^{\a}) .
\end{equation}

The general solution of Eq. \p{dep3}
with respect to lower indices
is  (cf. with \p{dep2s})
  \begin{equation}\label{dep3s}
   \Psi_{a \b } ^{~~ \a}  = {1 \over 10}
\s_{a~\b\g}
\tilde{\s}^{b \g \d}\Psi_{b \d}^{~~ \a}  , \qquad
\end{equation}

   To solve Eq. \p{dep3} we have to substitute the general decomposition
  \begin{equation}\label{Psidec}
   \Psi_{a\b} ^{~~ \a} \equiv
   \Psi_{0~a} \d_{\b } ^{~ \a}  + \Psi_{2~a}^{~~~cd} (\s_{cd})_{\b } ^{~ \a}
   +   \Psi_{4~a}^{~~~c_1...c_4} (\s_{c_1...c_4})_{\b } ^{~ \a}
\end{equation}
and use the $D=10$ sigma matrix algebra
 $$
\s^a \tilde{\s}^{c_1...c_q} =  \s^{ac_1...c_q} +
 q \eta^{a[c_1} \s^{c_2...c_q]},
 $$
 $$
 \s^{c_1...c_5} =  - {1 \over 5!} \e^{c_1...c_5d_1...d_5}
 \s_{d_1...d_5}, \qquad
 \s^{c_1...c_6} =  - {1 \over 4!} \e^{c_1...c_6d_1...d_4}
 \s_{d_1...d_4}
$$
 to  decompose  Eq. \p{dep3s} onto the irreducible parts.
As a result, we obtain the general solution of \p{dep3s}
   \begin{equation}\label{dep4}
  \Psi_{a \b} ^{~~ \a} \equiv \Psi_{0~a}
   \d_{\b } ^{~ \a}  +
   (\eta_{a[c }\Psi_{0~d]}+
   4\Psi_{3~[acd]})
   (\s^{cd})_{\b } ^{~ \a} +
   (\Psi_{5~[ac_1...c_4]}^{[-]}
   + \eta_{a[c_1} \Psi_{3~c_2c_3c_4]})
    (\s^{c_1...c_4})_{\b } ^{~ \a} ,
   \end{equation}
  where the following notations for {\sl independent} irreducible parts of
  Eq. \p{J2} are used
  \begin{equation}\label{defPsi0}
  \Psi_{0~a} \equiv {1 \over 16}
  \Psi_{a \a} ^{~~ \a}
   \end{equation}
  \begin{equation}\label{defPsi3}
   \Psi_{3}^{[bcd]} \equiv
  \Psi_{a\b} ^{~~ \a}
    (\s^{abcd})_{\a } ^{~ \b},
   \end{equation}
  \begin{equation}\label{defPsi5}
   \Psi_{5~[ac_1...c_4]}^{[-]} \equiv
  \Psi_{b\b} ^{~~ \a}
    (\s_{b_1...b_4})_{\a } ^{~ \b} (\d^b_{[a} \d^{b_1}_{c_1} ...
    \d^{b_4}_{c_4]} - {1 \over 5!} \e_{ac_1...c_4}
    {}_{}^{bb_1...b_4}).
    \end{equation}
    Eqs. \p{redef1},
   \p{defPsi0} --\p{defPsi5} defines the parts of Eq. \p{I2}
   which {\sl can be }
   independent on Eq.  \p{I1}.

   \bigskip

   \bigskip

 \subsection*{B2. Gauge field constraints and fermionic superembedding
 condition}

 Here we use the established dependence of the higher order components of
 the integrability conditions \p{JFF} to obtain a relation of
 \p{JFF} with the fermionic superembedding equation \p{E2hE11}.

\bigskip

 The complete explicit form of Eq. \p{JFF} is given by
\begin{equation}\label{JFF1}
J_3 \equiv
d({\cal F} - F)= -H_3 - E^b \wedge T^a F_{ab}
- {1 \over 2} E^b \wedge E^a \wedge d F_{ab} =
\qquad
\end{equation}
$$
- i  E^b \wedge E^{1\a} \wedge E^{1\b} \s^a_{\a\b} (\eta - F)_{ab} +
 i  E^b \wedge E^{2\a} \wedge E^{2\b} \s^a_{\a\b} (\eta + F)_{ab}
- {1 \over 2} E^b \wedge E^a \wedge d F_{ab} .
$$

If one uses the induced supervielbeine \p{ea2}, \p{eal2}  to decompose
the world-volume superspace differential  \p{Dind} ${\cal D}= E^a D_a +
E^{1\a} D_{\a}$ and extracts the expression
\begin{equation}\label{calE}
{\cal E}^{\a} \equiv E^{2\a} - E^{1\b} h_\b^{~\a} - E^a \psi_a^\a
\equiv  E^{1\b} (E_\b^{2\a} - h_\b^{~\a}), \qquad
\end{equation}
$$
  \psi_a^{~\a} = D_a \Theta^{2\mu} \d_\mu^{~\a}
$$
from all the terms in the r.h.s. of Eq. \p{JFF1}
(i.e. substituting here
$E^{2\a} = {\cal E}^{\a} + E^{1\b} h_\b^{~\a} + E^c \psi_c^{~\a}$),
one gets
\begin{equation}\label{JFF2}
J_3 \equiv
d({\cal F} - F)=
- i  E^b \wedge E^{1\a} \wedge E^{1\b}
(\s^a_{\a\b} (\eta - F)_{ab} -
 (h\s^a h^T)_{\a\b} (\eta + F)_{ab}) -
\end{equation}
$$
- {1 \over 2}
E^b \wedge E^c \wedge E^{1\a} J_{\a c b}
- {1 \over 2}
E^b \wedge E^c \wedge E^d J_{dcb} +
$$
$$
+ i E^b \wedge (E^{2\a} + E^{1\b} h_\b^{~\a} + E^a \psi_a^\a)
\wedge
{\cal E}^{\b}
 \s^a_{\a\b} (\eta + F)_{ab}
$$
where $J_{\a cb},~J_{bcd}$ are defined by  Eqs.
\p{J2} and \p{J3}
\begin{equation}\label{J21}
J_{\a b c}
= D_{\a}F_{bc} + 4i (h\s^a\psi_{[b})_{\a} (\eta - F)_{c]a} ,
\qquad
\end{equation}
\begin{equation}\label{J31}
J_{abc}=
D_{[a } F_{bc]} + 2i( \eta + F)_{d[a}  (\psi_{b}\s^{d} \psi_{c]}) .
\qquad
\end{equation}

Taking into account the fermionic superembedding equations
\p{E2hE11} following from the generalized action, one obtains
the set of integrability conditions \p{J1}, \p{J2}, \p{J3}
considered in the previous section.

As it was already proved, such integrability conditions can be used to
obtain the fact that the spin tensor field $h$ is Lorentz group valued
\p{Id1},  \p{hinS}
as well as the relation of $h$ superfield with the gauge field strength
\p{Id2}.

On the other hand, if we suppose that Eqs. \p{Id1}, \p{Id2}
 are satisfied, the first term in \p{JFF2} vanishes identically.
 The conditions of such vanishing coincides with Eq. \p{J1}.
 Henceforth, in accordance with dependence relations being established
 in the previous Subsection, the second and third terms in \p{JFF}
 (given by the Eqs. \p{J21} and \p{J31}) vanish as a result of
 \p{J1}.

 Thus, in the assumption that Eqs. \p{Id1}, \p{Id2}
 holds, the integrability  condition \p{JFF} for the world volume gauge
 field constraints \p{dAF} acquires the form
\begin{equation}\label{JFF3}
J_3 \equiv
d({\cal F} - F)=
 i
E^b
\wedge
(E^{2\a} + E^{1\b} h_\b^{~\a} + E^c \psi_c^\a)
\wedge
{\cal E}^{\b}
 \s^a_{\a\b} (\eta + F)_{ab} = 0,
\end{equation}

 Decomposing Eq. \p{JFF3} onto the basic 3-forms of the world volume
 superspace, one can find that
 it contains two equations

 \begin{equation}\label{JFF31}
(E_\a^{2\a^\prime}- h_\a^{~\a^\prime})
\s^a_{\a^\prime \b^\prime}
(E_\b^{2\b^\prime}+ h_\b^{~\b^\prime}) +
(E_\b^{2\a^\prime}- h_\b^{~\a^\prime})
\s^a_{\a^\prime \b^\prime}
(E_\a^{2\b^\prime}+ h_\a^{~\b^\prime}) = 0
\end{equation}
 \begin{equation}\label{JFF32}
(E_\a^{2\a^\prime}- h_\a^{~\a^\prime})
\s^a_{\a^\prime \b} \psi_{[c|}^{~\b} (\eta +F)_{a |b]}
= 0.
\end{equation}

Eq. \p{JFF31} can be simplified to the form
 \begin{equation}\label{JFF311}
E_\a^{2\a^\prime}
\s^a_{\a^\prime \b^\prime}
E_\b^{2\b^\prime}
= h_\a^{~\a^\prime}
\s^a_{\a^\prime \b^\prime} h_\b^{~\b^\prime} \equiv
\s^b_{\a\b} ((\eta - F)(\eta + F)^{-1})_b^{~a} .
\end{equation}

As it was noted above,
the second equation in \p{JFF311} defines
$h$ spin-tensor completely up to a sing.
Hence, we can conclude that Eq. \p{JFF311} has two solutions
 \begin{equation}\label{JFF311+}
E_\a^{2\b} =  h_\a^{~\b} ,
\end{equation}
 \begin{equation}\label{JFF311-}
E_\a^{2\b} = - h_\a^{~\b} .
\end{equation}

For the first solution, the second equation  \p{JFF32} is satisfied
identically,
while for the second solution Eq. \p{JFF32}
becomes
 \begin{equation}\label{JFF32-}
h_\a^{~\a^\prime}
\s^a_{\a^\prime \b} \psi_{[c|}^{~\b} (\eta +F)_{a |b]}
\equiv
h_\a^{~\a^\prime}
\s^a_{\a^\prime \b} \psi_{[c}^{~\b} (\eta -F)_{b]a}
= 0,
\end{equation}
and provides a nontrivial additional restriction for the
bosonic derivative of the  $\Theta^2$ superfield
$\psi_a^{~\a}=D_a\Theta^{2\mu}\d_\mu^{~\a}$.

Comparing \p{JFF32-} with (the dependent) Eq. \p{J2} one can find that
\p{JFF32-} signifies
$
D_\a F_{ab} = 0
$
and, hence, remove the world volume gauge  field strength from the
consideration. Thus only the solution \p{JFF311+} is nontrivial.

\bigskip

Generalized action \p{SLLL} selects the solution
 \p{JFF311+} uniquely.

%
%

\end{document}